\documentclass[a4paper,11pt]{article}
\usepackage{jheppub} % for details on the use of the package, please see the JINST-author-manual
\usepackage{lineno}
\usepackage[dvipsnames]{xcolor} 
\usepackage{dirtytalk}
\usepackage{ytableau}
\usepackage[dvipsnames]{xcolor}
\usepackage{dsfont}
\usepackage{mathdots}
\usepackage{colortbl}
\usepackage{braket}
\usepackage{amsthm}
\usepackage{amsmath}
\usepackage{float}
\usepackage{bbm, dsfont}
\usepackage[normalem]{ulem}
\usepackage{slashed}
\usepackage{mathtools}
\usepackage{multirow}
\usepackage{cancel}
\usepackage{cleveref}
\usepackage{orcidlink}
\usepackage{enumerate}
\usepackage{fontawesome}
\usepackage{adjustbox}
\usepackage{tcolorbox}
\usepackage{relsize}
\usepackage{csquotes}
\usepackage{mathrsfs} 
\usepackage{wasysym}

\DeclareMathOperator{\sgn}{sgn}

\usepackage{tikz}
    \usetikzlibrary{positioning}
    \usetikzlibrary{arrows.meta}
    \usetikzlibrary{angles,quotes}
    \usetikzlibrary{arrows}
    \usetikzlibrary{shapes}
    \usepackage{pgfplots}
    \usetikzlibrary{calc}
    \usetikzlibrary{decorations.markings}
     \usetikzlibrary{decorations.pathmorphing}
    \tikzset{snake it/.style={decorate, decoration=snake}}

    \usetikzlibrary{matrix}

    \pgfplotsset{compat=1.18}

\definecolor{ourblue}{RGB}{0, 57, 120} % 
\colorlet{ourlightblue}{ourblue!50!white}
\newcommand{\eps}{\varepsilon}
\newcommand{\rd}{\mathrm{d}}
\newcommand{\cS}{\mathcal{S}}
\newcommand{\cH}{\mathcal{H}}
\newcommand{\cM}{\mathcal{M}}
\newcommand{\cQ}{\mathcal{Q}}
\DeclareMathOperator{\Gram}{Gram}
\DeclareMathOperator{\Vol}{Vol}
\newcommand{\iu}{{i\mkern1mu}}
\DeclareMathOperator{\SV}{SV}
\DeclareMathOperator{\id}{id}
\DeclareMathOperator{\Li}{Li}
\DeclareMathOperator{\ALi}{ALi}

\DeclareMathOperator{\perR}{per_\mathbb{R}}
\newcommand{\Df}{\mathrm{d}}
\newcommand{\Q}{\mathcal{Q}}

\makeatletter
\newcommand*{\rom}[1]{\expandafter\@slowromancap\romannumeral #1@}
\makeatother

\def\beq{\begin{equation}}
\def\eeq{\end{equation}}
\def\bsp#1\esp{\begin{split}#1\end{split}}

\author{Claude Duhr${}^1$, Paul Mork${}^1$}
\affiliation{
\vskip 0.5 em
${}^1$Bethe Center for Theoretical Physics, Universit\"at Bonn, D-53115, Germany\\
\vskip 0.5 em}

\emailAdd{cduhr@uni-bonn.de, pmork@uni-bonn.de}

\title{Analytic results for one-loop integrals in dimensional regularisation}

\abstract{We present a method to obtain analytic results in terms of multiple polylogarithms for one-loop triangle, box and pentagon integrals depending on an arbitrary number of scales and to any desired order in the Laurent expansion in the dimensional regulator $\varepsilon$. Our method leverages the fact that for $\varepsilon=0$ one-loop integrals compute volumes of simplices in hyperbolic spaces, which can always be evaluated in terms of polylogarithms using an algorithm recently introduced in pure mathematics. The higher orders in $\varepsilon$ can then be expressed as a one-fold integral involving the result for $\varepsilon=0$. Remarkably, we find that for up to five external legs, all integrals can be evaluated algorithmically in terms of polylogarithms using direct integration techniques, which, in particular, requires us to rationalise all appearing square roots. We also discuss how we can use the connection to hyperbolic geometry to perform the analytic continuation from the Euclidean region to other kinematic regions.}

\begin{document}
\begin{flushright}
    BONN-TH/2025-34
    \end{flushright}

\maketitle
\flushbottom

% !TEX root = main.tex

\section{Introduction}
\label{sec:introduction}

The computation of higher-loop corrections to scattering amplitudes is one of the main bottlenecks when it comes to making precise predictions for collider experiments (see,~e.g., ref.~\cite{Caola:2022ayt} for a recent review). A notable exception is the set of one-loop integrals required for next-to-leading order (NLO) computations. It is well known that all one-loop scattering amplitudes in four space-time dimensions can be expressed in terms of scalar integrals with at most four external legs, up to terms that vanish in four dimensions. The relevant integrals have been evaluated analytically a long time ago~\cite{tHooft:1978jhc}, and there are implementations into public computer codes, see,~e.g.,~refs.~\cite{Ellis:2007qk,Carrazza:2016gav}. A distinctive feature of these analytic results in four space-time dimensions is that they can be entirely expressed in terms of logarithm and dilogarithm functions evaluated at algebraic arguments.

The previous discussion may give the impression that the computation of one-loop integrals is a solved problem. However, this is not so. At higher orders in perturbation theory, also products of lower-loop integrals need to be considered, as well as products of lower-loop amplitudes multiplied by ultraviolet or infrared counterterms. When working in dimensional regularisation~\cite{tHooft:1972tcz,Cicuta:1972jf,Bollini:1972ui}, this implies that one also needs to consider higher orders in the $\eps$-expansion of one-loop integrals, because they can contribute to the finite terms of the amplitude upon multiplication with a pole in $\eps$.\footnote{Though it is an open problem if one can devise a strategy such that these contributions explicitly cancel in the end, cf.,~e.g.,~ ref.~\cite{Weinzierl:2011uz}.} When working in a  `t Hooft-Veltman-like scheme, where the external momenta are taken to lie in four dimensions and only the loop momenta are allowed to access the $(D-4)$ remaining dimensions, one needs to consider up to pentagon integrals in dimensional regularisation, including the higher orders in the dimensional regulator $\eps$. However, only very few results are known for scalar one-loop integrals to higher orders in the dimensional regulator (see,~e.g.,~refs.~\cite{DelDuca:2009ac,Dixon:2011ng,DelDuca:2011ne,DelDuca:2011jm,DelDuca:2011wh,Chavez:2012kn,Kozlov:2015kol,Bourjaily:2019exo,Buccioni:2023okz,Becchetti:2025osw}), even when restricted to at most five external legs. This can be traced back to the fact that the higher orders in $\eps$ will typically involve more complicated transcendental functions than the logarithm and dilogarithm functions that were sufficient for NLO computations.

Via dimensional shift identities~\cite{Tarasov:1996bz,babis_thesis,Lee:2009dh}, the computation of one-loop integrals to higher orders in $\eps$ is closely related to evaluating these integrals in integer dimensions different from four. It is well known that one-loop integrals in integer space-time dimensions are connected to volumes of simplices in hyperbolic spaces~\cite{Davydychev:1997wa,Schnetz:2010pd,Mason:2010pg,Nandan:2013ip}. Volumes of hyperbolic simplices are a well-studied subject in pure mathematics. In particular, they are governed by Schl\"afli's differential equation~\cite{Schlafli1901}, and in hyperbolic three- or five space they can always be expressed in terms of dilogarithm or trilogarithm functions, respectively~\cite{Zagier1991Dilogarithm,Kellerhals1992Orthoschemes,Kellerhals1995Volumes}. Very recently, based on earlier work in ref.~\cite{Goncharov1999Volumes}, a general algorithm to evaluate volumes of simplices in hyperbolic spaces in terms of multiple polylogarithms~\cite{Goncharov:1998kja} was presented~\cite{Rudenko2020Orthoschemes}. In ref.~\cite{Ren:2023tuj} this result was used to obtain an algorithm to evaluate general one-loop integrals in integer space-time dimensions in terms of multiple polylogarithms (previously, the connection to hyperbolic simplices had been used to compute the symbols~\cite{Goncharov:2010jf} of certain classes of one-loop integrals~\cite{Spradlin:2011wp}). However, the relationship between one-loop integrals and volumes in hyperbolic spaces is restricted to integrals in integer space-time dimensions, and it is currently not known if or how it extends to higher orders in the Laurent expansion in the dimensional regulator.

The goal of this paper is to leverage the recent advances in how to compute analytic results for one-loop integrals in integer space-time dimensions to compute higher orders in the dimensional regulator for triangle, box and pentagon integrals. The main idea is to use the fact that in a `t Hooft-Veltman-like scheme, the higher orders in $\eps$ can be obtained by integrating the result for $\eps=0$ over the propagator masses. The result for $\eps=0$ can in turn be expressed in terms of multiple polylogarithms using the connection to volumes of hyperbolic simplices, in particular the results of ref.~\cite{Ren:2023tuj}. We obtain in this way a one-fold integral representation of the higher orders in the dimensional regulator, and one may attempt to perform these integrations order by order in $\eps$ using direct integration techniques~\cite{Brown2009MasslessHigherLoop,Anastasiou:2013srw,Ablinger:2014yaa,Panzer:2014caa,hyperlogprocedures,Kardos:2025klp}. At this point, however, one faces an obstacle: the arguments of the multiple polylogarithms obtained via the method of ref.~\cite{Ren:2023tuj} involve square roots, and direct integration techniques are only applicable if all these square roots can be rationalised. Remarkably, we find that we can always rationalise all the square roots for triangle, box and pentagon integrals. We obtain in this way an algorithm to compute triangle, box and pentagon integrals for arbitrary mass configurations, and thus for all one-loop integrals relevant in a `t Hooft-Veltman-like scheme.

Our paper is organised as follows: in section~\ref{sec:setup} we briefly review one-loop integrals and their connection to volumes in hyperbolic spaces. In section~\ref{sec:dimreg} we discuss how we can leverage the results for the hyperbolic volumes to obtain integral representations for the higher orders in the $\eps$-expansion by integrating over an auxiliary mass parameter. In section~\ref{sec:examples} we illustrate how we can use this approach to obtain explicit results for triangle, box and pentagon integrals depending on many scales in the Euclidean region, and we discuss the analytic continuation to other regions in section~\ref{sec:analytic_continuation}. In section~\ref{sec:conclusions} we draw our conclusions. We include appendices where we present some technical proofs omitted throughout the main text and ancillary files with our analytic results.

% !TEX root = main.tex

\section{One-loop integrals and volumes in hyperbolic space}
\label{sec:setup}

\subsection{Generalities}
The main focus of our paper are $N$-point one-loop integrals depending on $N$ external momenta $p_i$ subject to momentum conservation $\sum_{i=1}^Np_i=0$. They may be defined as
\begin{equation}\label{eq:integral_def}
    I^D_N (\{p_i\}, \{m_j\}) = e^{\gamma_E\eps}\int \frac{\mathrm{d}^D l}{\pi^{D/2}} \prod_{i=1}^N \frac{1}{\left(l - l_i\right)^2 + m_i^2}\,,
\end{equation}
where $\gamma_E=-\Gamma'(1)$ is the Euler-Mascheroni constant, the $m_i$ are the propagator masses, and the momenta $l_i$  denote partial sums of external momenta, i.e., $l_i = \sum_{j=1}^{i-1} p_j$. Alternatively, we can see the $l_i$ as defining points in a dual position space~\cite{Drummond:2006rz}. We work in dimensional regularisation~\cite{tHooft:1972tcz,Cicuta:1972jf,Bollini:1972ui} in $D=d-2\eps$ dimensions, where $d$ is an even integer. We are typically only interested in the Laurent expansion of the integrals around $\eps=0$. Note that the expression in eq.~\eqref{eq:integral_def} is obtained after Wick rotation from Minkowski to Euclidean space, as visible from the sign of the squared mass in the denominators. 

The integrals in eq.~\eqref{eq:integral_def} are sufficient to cover all one-loop integrals, because they can all be expressed in terms of the latter via integration-by-parts (IBP) relations~\cite{Tkachov:1981wb,Chetyrkin:1981qh}  and/or tensor reduction~\cite{tHooft:1978jhc}. In particular, we do not need to consider integrals with numerators and/or raised to higher propagator powers. Moreover, dimensional recurrence relations relate integrals in $D$ and $D\pm2$ dimensions~\cite{Tarasov:1996bz,babis_thesis,Lee:2009dh}. We can then always choose a set of basis integrals where $d=N$ or $N+1$. Finally, if we restrict ourselves to the computation of scattering amplitudes in $d=4$ dimensions and we neglect contributions that vanish as the dimensional regulator $\eps$ tends to zero, then one only needs to consider integrals with $N\le 4$~\cite{Melrose:1965kb}.

Analytic results are available for all one-loop integrals in $D=4-2\eps$ dimensions up to terms that vanish as $\eps\to0$, cf.~e.g., refs.~\cite{tHooft:1972tcz,Ellis:2007qk,Carrazza:2016gav}. In some cases closed expressions valid to all orders in $\eps$ in terms of hypergeometric functions exist, cf.,~e.g.,~refs.~\cite{Davydychev:1990jt,Gehrmann:1999as,Fleischer:2003rm,Brandhuber:2007yx,Kniehl:2010aj,davydychev_four-point_2018}
%\claudecomment{Here we should check if we can find more!}\paulcomment{Maybe \cite{Davydychev:1990jt,davydychev_four-point_2018} works of Davydychev using their splitting procedure to split into Feynman integrals which can be evaluated as 2F1 and Lauricella for triangle box}. 
The situation, however, is very different if one considers also higher-order terms in the dimensional regulator and/or integrals with more than four propagators in different space-time dimensions. Only very few analytic results of this type are available, see for example refs.~\cite{DelDuca:2009ac,Dixon:2011ng,DelDuca:2011ne,DelDuca:2011jm,DelDuca:2011wh,Chavez:2012kn,Dixon:2013eka,Kozlov:2015kol,Kozlov:2016vqy,Gehrmann:2018yef,Bourjaily:2019exo,Henn:2022ydo,Buccioni:2023okz,Becchetti:2025osw}.
It is expected that all the coefficients of the Laurent expansion in $\eps$ of one-loop integrals can always be expressed in terms of multiple polylogarithms~\cite{Goncharov:1998kja} (see eq.~\eqref{eq:MPL_def} for the definition) and the symbols of all Laurent coefficients are known conjecturally~\cite{Abreu:2017enx,Abreu:2017mtm} (though, to the best of our knowledge, a formal mathematical proof is currently still lacking).

A notable exception are integrals for $D=d$ (i.e., for $\eps=0$). It has been shown a long time ago that these integrals compute volumes of polytopes in spherical or hyperbolic spaces~\cite{Davydychev:1997wa,Schnetz:2010pd,Mason:2010pg,Nandan:2013ip}.  Using a connection between the theory of motives and volumes in hyperbolic spaces~\cite{Goncharov1999Volumes}, the symbols~\cite{Goncharov:2010jf} of all one-loop integrals with $D=N$ were obtained~\cite{Spradlin:2011wp,Arkani-Hamed:2017ahv}. If convergent, these integrals are dual conformally invariant. It is then possible to write integrals with $D=N+1$ as limits of dual conformally invariant integrals, with one of the points sent to infinity. Note that the conformal invariance, and even a stronger Yangian invariance, carries over to integrals with massive propagators~\cite{Henn:2011by,Loebbert:2020hxk,Loebbert:2020glj}. Very recently, using some novel insight from pure mathematics~\cite{Rudenko2020Orthoschemes}, it was possible to obtain analytic expressions in terms of polylogarithms for all one-loop integrals with $D=N$ or $N+1$~\cite{Ren:2023tuj}. The connection between one-loop integrals and volumes of polytopes in hyperbolic space, however, is restricted to integrals in integer dimension, and it is not clear if and how this connection extends to dimensional regularisation. The aim of this paper is to leverage the connection between one-loop integrals and volumes in hyperbolic space to obtain an algorithm to compute the higher orders in the Laurent expansion in $\eps$ for one-loop integrals with $N\le 5$. Before we discuss our method and our results, we review in the remainder of this section the connection between one-loop integrals in integer dimensions and volumes in hyperbolic spaces.

\subsection{One-loop integrals in integer dimensions as volumes in hyperbolic space}
\label{sec:hyperbolic}

%\paragraph{The case $D=N$.}
We start the discussion by focusing on the case $D=N$, and we closely follow refs.~\cite{Davydychev:1997wa,Ren:2023tuj}. We first review some basic notions of hyperbolic spaces.
Consider the hyperboloid in Euclidean $N$-space defined by the equation
\beq
-\alpha_N^2+\sum_{k=1}^{N-1}\alpha_k^2 = -1\,.
\eeq
Points on this hyperboloid have the form $\alpha:=\Big(\alpha_1,\ldots,\alpha_{N-1},\pm\sqrt{1+\sum_{k=1}^{N-1}\alpha_k^2}\Big)$, and consequently the hyperboloid has two branches. In the following we focus on the branch where the square root is positive. For our purposes another representation of this hyperboloid -- called the \emph{projective model} -- is obtained by assigning to a point $\alpha$ on the positive branch the point $p$ inside the $(N-1)$-dimensional unit ball defined by
\beq
p =(p_1,\ldots,p_{N-1})= \frac{1}{\sqrt{1+\sum_{k=1}^{N-1}\alpha_k^2}}\,(\alpha_1,\ldots,\alpha_{N-1})\,.
\eeq
The scalar product in the projective model is
\beq
\langle p,q\rangle := 1-\sum_{k=1}^{N-1}p_kq_k\,.
\eeq
Both the projective model, i.e., the unit $(N-1)$-ball, and the positive branch of the hyperboloid are models for the hyperbolic $(N-1)$-space $\mathbb{H}^{N-1}$. Points $p$ on the boundary of the unit ball (which is the $(N-2)$-dimensional unit sphere $S^{N-2}$) have norm $\langle p,p\rangle=0$ and correspond to the points at infinity of hyperbolic space.

Consider now $N$ points $v_1,\ldots,v_N$ in $\mathbb{H}^{N-1}$. If we connect these points by hyperbolic geodesics, we obtain a simplex $\cS$ in $\mathbb{H}^{N-1}$ with vertices $v_k$. We define the \emph{Gram matrix} of this simplex as the matrix $\Gram(\cS) := \big(\langle v_i, v_j\rangle)_{1\le i,j\le N}$. A general polytope with more vertices can always be dissected into a union of simplices. It is possible to show that the volume of the simplex $\cS$ is given by
\beq\label{eq:volume_general}
\Vol(\cS) = \sqrt{|\det \Gram(\cS)|}\,  \int_0^\infty \frac{\delta(\alpha_N - 1)\,\prod_{i=1}^N \mathrm{d} \alpha_i}{\left(\sum_{i,j=1}^N  \langle v_i, v_j\rangle\,\alpha_i\, \alpha_j\right)^{N/2}}\,.
\eeq

We may now compare the expression for the volume of a hyperbolic simplex in eq.~\eqref{eq:volume_general} to the 
 Feynman-parameter representation of a one-loop integral with $D=N$. The latter can be cast in the form,
\begin{equation}\label{eq:I_N^N_integral}
    I_N^N (Q) = \Gamma\! \left( \tfrac{N}{2} \right) \int_0^\infty \frac{\delta(\alpha_N - 1)\,\prod_{i=1}^N \mathrm{d} \alpha_i }{\left(\sum_{i,j=1}^N Q_{i,j}\, \alpha_i\, \alpha_j\right)^{N/2}}\,,
\end{equation}
where all the kinematic dependence is encoded into the matrix %\claudecomment{Is this the modified Cayley matrix?}\paulcomment{no; the modified Cayley would be the extended matrix as far as I understand it}
\beq\label{eq:Cayley}
Q_{i,j} = \frac{(l_i - l_j)^2 + m_i^2 + m_j^2}{2}\,.
\eeq
Comparing eqs.~\eqref{eq:volume_general} and~\eqref{eq:I_N^N_integral}, we see that we have the relation~\cite{Davydychev:1997wa,Schnetz:2010pd}
\begin{equation}\label{eq:INN_Vol}
    I_N^N (Q) = \Gamma\! \left( \tfrac{N}{2} \right) \frac{\operatorname{Vol}(\cS_Q)}{\sqrt{|\det Q|}}\,,
\end{equation}
where $\cS_Q$ is a hyperbolic simplex with Gram matrix $Q$. In other words, we can find $N$ points $v_1,\ldots,v_N$ in $\mathbb{H}^{N-1}$ that span a simplex $\cS_Q$ with Gram matrix $\Gram(\cS_Q)=Q$, then the volume of $\cS_Q$ is proportional to the value of the Feynman integral $I_N^N(Q)$.

From eq.~\eqref{eq:INN_Vol} we see that the computation of $I_N^N (Q)$ (with $N$ even) is reduced to the evaluation of the volume of the simplex $\cS_Q$. While there is currently no known formula that computes $\operatorname{Vol}(\cS_Q)$ directly (though there is a general differential equation for it~\cite{Schlafli1901}), it is possible to dissect the simplex into a union of a special class of simplices, called \emph{orthoschemes}, for which volume formulas exist. One possible definition of an orthoscheme is that there is an ordering $(v_1,\ldots,v_n)$ on its sets of vertices such that the consecutive edges $(v_{i-1},v_i)$ and $(v_i,v_{i+1})$ are mutually orthogonal. Before we review how to compute volumes of orthoschemes in subsection~\ref{sec:orthoschemes}, we describe an algorithm for decomposing a general simplex into orthoschemes.

\subsection{Dissecting simplices into orthoschemes (1)}\label{sec:dissection1}
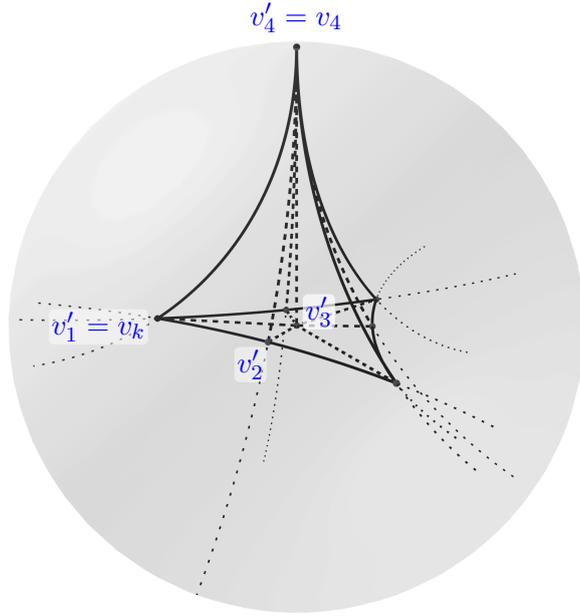
\begin{figure}[!th]
    \centering
    \begin{tikzpicture}
        \node[inner sep=0pt] (img) {%
            \includegraphics[width=0.5\linewidth]{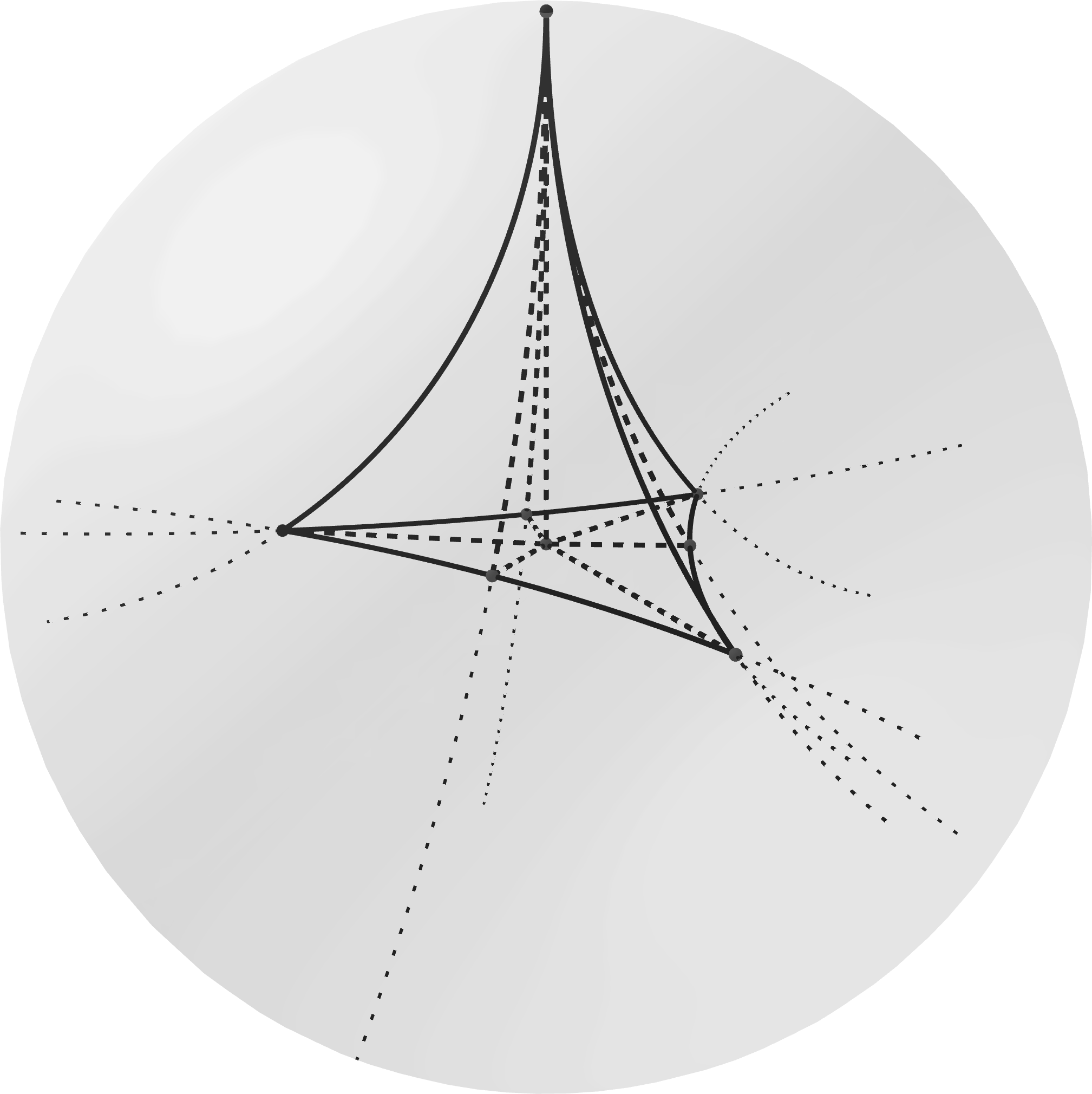}
        };
        \node[text=blue,fill=white,fill opacity=0.5,text opacity=1,rounded corners=2pt,inner sep=1pt] at (0.0,4.1) {$v'_{4}=v_4$};
        \node[text=blue,fill=white,fill opacity=0.5,text opacity=1,rounded corners=2pt,inner sep=1pt] at (-2.6,0.0) {$v'_{1}=v_k$};
        \node[text=blue,fill=white,fill opacity=0.5,text opacity=1,rounded corners=2pt,inner sep=1pt] at (-0.6,-0.5) {$v'_{2}$};
        \node[text=blue,fill=white,fill opacity=0.7,text opacity=1,rounded corners=2pt,inner sep=1pt] at (0.3,0.2) {$v'_{3}$};
    \end{tikzpicture}
    \caption{Dissecting a three-simplex into six orthoschemes in the Poincaré ball model.}
    \label{fig:dissection_N}
\end{figure}

The decomposition of a simplex in $\mathbb{H}^{N-1}$ into a union of orthoschemes is not unique.
An algorithm to achieve a decomposition was presented in ref.~\cite{Ren:2023tuj}. It proceeds recursively and results in a decomposition of the simplex $\cS_Q$  into \((N-1)!\) orthoschemes. In detail, the algorithm of ref.~\cite{Ren:2023tuj} proceeds via the following steps:
\begin{enumerate}
    \item Choose a vertex \(v_N\) of the simplex \(\cS\).
    \item Project \(v_N\) orthogonally onto the codimension-one face \(\mathcal{F}_N\) that does not contain \(v_N\). This projection yields a point \(v'_{N-1}\). Note that \(\mathcal{F}_N\) is itself a simplex of dimension \(N-2\).
    \item Project \(v'_{N-1}\) orthogonally onto all codimension-one faces \(\mathcal{F}_{N,j_1}\) of \(\mathcal{F}_{N}\), $j_1=1,\ldots,N-1$. These projections yield points \(v'_{N-2,j_1}\) lying on the corresponding faces \(\mathcal{F}_{N,j_1}\).
    \item Proceed recursively, i.e., continue the projection process until the faces \(\mathcal{F}_{N,j_1,\ldots,j_{N-2}}\) become zero-dimensional, i.e., \(v_{1,j_{N-2}} = v_k\) is one of the vertices (except \(v_N\)) of \(\cS\).
    \item The resulting convex hulls $\operatorname{conv}(v'_N, v'_{N-1}, v'_{N-2,j_1}, \ldots, v'_{1,j_{N-2}})$, with \(v'_N = v_N\) and \(v'_{1,j_{N-2}} = v_k\) for some \(k \neq N\), are orthoschemes.
\end{enumerate}
We denote the set of orthoschemes obtained from this procedure as \(\operatorname{Dis}_{v_N}(Q)\).
An example of this dissection procedure for $N=4$ is shown in figure~\ref{fig:dissection_N}.
In the following, we represent orthoschemes and simplices simply by their Gram matrix $\cQ$. For example, we write $\operatorname{Vol}(\cQ)$ instead of $\operatorname{Vol}(\cS_\cQ)$, etc. 
%This type of dissection was studied in \cite{Ren:2023tuj}; detailed expressions can be found there.
At the end of the dissection procedure, we obtain an expression for the volume, and thus of the Feynman integral, as a sum of volumes of orthoschemes,
\begin{equation}\label{eq:diss_Ren}
    I^N_N (Q) = \Gamma\! \left( \tfrac{N}{2} \right) \frac{1}{\sqrt{| \det Q | }} \sum_{\cQ\in \operatorname{Dis}_{v_N} \!(Q)} \operatorname{sgn}(\cQ)\operatorname{Vol}(\cQ)\,.
\end{equation}
%The sum in eq.~\eqref{eq:diss_Ren} runs over all orthoschemes with Gram matrix $\cQ$ appearing in the decomposition.
The sign factor $\operatorname{sgn}(\cQ)$ is introduced to account for the fact that that vertices \(v'_i\) might lie outside of the face they were projected to. It will be discussed in the next section.

%Since this procedure does not always provide a proper dissection, we have to insert a potential sign factor \(\operatorname{sgn} (\cS_\cQ)\), which compensates for the fact that vertices \(v'_i\) might lie outside of the face they were projected to.

%\begin{figure}[!th]
%    \centering
%    \includegraphics[width=0.5\linewidth]{res/triangle_ortho_fix_crop.png}
%    %\vspace{96pt}
%    %\incfig[.4]{3-simplex_Diss4}
%    \caption{Dissecting a three-simplex into six orthoschemes in the Poincar\'e ball model}
%    \label{fig:dissection_N}
%\end{figure}

\subsection{Volumes of orthoschemes}
\label{sec:orthoschemes}
At this point, we have reduced the computation of $I_N^N(Q)$ with $N$ even to the computation of volumes of orthoschemes in $\mathbb{H}^{N-1}$. 
For $\mathbb{H}^3$, these are known to be expressible in terms of the dilogarithm function (see, e.g., ref.~\cite{Zagier1991Dilogarithm}), while for $\mathbb{H}^5$ they can be expressed in terms of trilogarithms~\cite{Kellerhals1992Orthoschemes,Kellerhals1995Volumes}. An algorithm to compute volumes of orthoschemes in $\mathbb{H}^{N-1}$ for general even values of $N$ in terms of multiple polyogarithms (MPLs)~\cite{Goncharov:1998kja} has recently been presented in ref.~\cite{Rudenko2020Orthoschemes}, building on earlier work in ref.~\cite{Goncharov1999Volumes}. 
MPLs can be defined recursively via the iterated integrals
\beq\label{eq:MPL_def}
G(a_1,\ldots,a_n;x) = \int_0^x\frac{\rd t}{t-a_1}G(a_2,\ldots,a_n;t)\,.
\eeq
The previous definition is valid whenever $a_n\neq0$. Otherwise we need to use the definition
\beq
G(\underbrace{0,\ldots,0}_{n};x) = \frac{1}{n!}\log^nx\,.
\eeq
The number $n$ of integrations is called the \emph{weight} of the MPL. We denote the vector space generated by MPLs of weight $n$ by $\cH_n$. MPLs enjoy many interesting properties. In particular, they form a (graded) Hopf algebra, equipped with an antipode $\operatorname{S} : \mathcal{H}_n \to \mathcal{H}_n$~\cite{Goncharov:1998kja,Brown2012MixedTate,Brown2017MotivicPeriods} (see also refs.~\cite{Duhr:2012fh,Duhr:2014woa} for a discussion in the context of Feynman integrals).\footnote{Strictly speaking, only the de Rham analogues of MPLs form a Hopf algebra. We will not dwell on this distinction, because for MPLs there is a natural projection from ordinary (motivic) MPLs to their de Rham analogues~\cite{Brown2017MotivicPeriods}. Since only single-valued analogues of MPLs appear in the computation of the volumes, this subtlety will not play a role here.}
Below we will mostly work with the representation of MPLs as nested sums,
\beq\label{eq:Li_def}
\Li_{m_1,\ldots,m_k}\!(z_1,\ldots,z_k) = \sum_{0<n_1<\cdots<n_k}\frac{z_1^{n_1}}{n_1^{m_1}}\cdots\frac{z_k^{n_k}}{n_k^{m_k}}\,.
\eeq
The number $k$ of sums is called the \emph{depth} of the MPL. The sum and integral representations are related by
\beq
\Li_{m_1,\ldots,m_k}\!(z_1,\ldots,z_k) = (-1)^k\,G(\underbrace{0,\ldots,0}_{m_k-1},\tfrac{1}{z_k},\ldots,\underbrace{0,\ldots,0}_{m_1-1},\tfrac{1}{z_1\cdots z_k};1)\,.
\eeq
MPLs are typically multivalued functions. We can associate to each MPL a single-valued version, which is real-analytic rather than holomorphic, satisfies the same holomorphic differential equation, and in addition it is free of branch cuts, i.e., it is single-valued. We denote by $\SV$ the map that assigns to an MPL its single-valued version.\footnote{For general periods, the map $\SV$ is only defined for de Rham periods~\cite{Brown2017MotivicPeriods}. However, by composing with the de Rham projection, we obtain a map that assigns a single-valued version to ordinary (motivic) MPLs.} Explicit constructions of the map $\SV$ can be found in refs.~\cite{BrownSVHPLs,brownSV,Brown2014SingleValued,Brown:2018omk,Dixon:2012yy,DelDuca:2016lad}.
%We briefly review this algorithm in this subsection.

In ref.~\cite{Goncharov1999Volumes} it was shown that the volume  of an \((N-1)\)-dimensional 
%\claudecomment{Here and below: should this be $n$ or $N$?} 
simplex \(\cS\) can be expressed via the \emph{real period map} $\operatorname{per}_{\mathbb{R}}$:
\begin{equation}
\label{eq:Vol_to_SV}
\operatorname{Vol}(\cS)=\frac{(-\iu)^{\frac{N}{2}+1}}{2^{\frac{N}{2}-1}\, \Gamma\! \left(\frac{N}{2} \right) }\, \operatorname{per}_{\mathbb{R}}(h(\cS))\,. % Put normalization factors like 2^n/2 ... in here instead of ortsP.
\end{equation}
Here $h(\cS)$ can be expressed in terms of multiple polylogarithms, and an operational definition sufficient for our purposes will be given below.
The map \( \operatorname{per}_{\mathbb{R}}: \mathcal{H}_n \longrightarrow \mathbb{R}(n-1) = \mathbb{R}(2\pi \iu)^{n-1} \) is known as the real period map~\cite{Goncharov1999Volumes}. An explicit construction of the real period map with examples is given in ref.~\cite{Ren:2023tuj}.
The real period map acting on $X\in\cH_n$ is related to the map $\SV$ through the formula \cite{Ren:2023tuj}:
\beq\bsp
    \operatorname{per}_\mathbb{R}(X) &= \frac{1}{4}\operatorname{SV} (X- \operatorname{S}(X)),\label{eq:realperSV}\\
    &= \begin{cases}
        \frac{1}{2}\, \Re (\SV(X)), & \textnormal{ if the weight of $X$ is odd}\,,\\
        \frac{\iu}{2}\,  \Im (\SV(X)), & \textnormal{ if the weight of $X$ is even}\,,
    \end{cases}
\esp\eeq
where \( \operatorname{S}\) is the antipode introduced earlier.

Let us now discuss how we can obtain the quantity $h(\cS)$ from eq.~\eqref{eq:Vol_to_SV}. Since we know how to dissect any simplex into a union of orthoschemes, it is sufficient to discuss $h(\cS)$ for $\cS$ being an orthoscheme. This is one of the main results of ref.~\cite{Rudenko2020Orthoschemes}. The first important ingredient in the construction of ref.~\cite{Rudenko2020Orthoschemes} is a bijection between generic orthoschemes in $\mathbb{P}^{N-1}$ 
%\paulcomment{should we include the full definition here for what a generic projective orthoscheme is? Generic is rather technical like hyperplanes in general position and projective is that we can have different smooth quadric}\claudecomment{Yes, we can do that!} 
and points in the moduli space $\cM_{0,N+2}$ of Riemann spheres with $(N+2)$ marked points.
A generic projective orthoschemes can be defined as a collection of a smooth quadric \(Q\) and an ordered set of hyperplanes \(H=(H_1,\ldots,H_N)\) in \(\mathbb{P}^{N-1}\), such that the hyperplanes are in general position and \(H_i \perp H_j\) for \(|i-j|>1\). This definition of an orthoscheme is a slight generalization of the definition we gave earlier.
We denote the orthoscheme associated to the point $z\in\cM_{0,N+2}$ by $\cS_z$. A point in $\cM_{0,N+2}$ can be represented as an $(N+2)$-tuple $z=(z_0,\ldots,z_{N+1})\in\big(\mathbb{CP}^1\big)^{N+2}$, with $z_i\neq z_j$ and modulo the natural action of SL$(2,\mathbb{C})$ on the Riemann sphere by M\"obius transformations. The SL$(2,\mathbb{C})$-action allows one to fix three points to \(z_0=0, z_N =1,\) and \(z_{N+1} = \infty\). The map from orthoschemes represented by the Gram matrix $\cQ$ to $\cM_{0,N+2}$ is then given by
\begin{equation}\label{eq:z_i_to_Q}
    z_i = \frac{\cQ_{i,N}^2}{\cQ_{i,i}\ \cQ_{N,N}}\,,\qquad 1\le i\le N-1\,.
\end{equation}
Using the bijective relationship between orthoschemes and the moduli space of Riemann spheres with marked points, it was shown in ref.~\cite{Rudenko2020Orthoschemes} that for an orthoscheme $\cS_z$ the function $h\left(\cS_z\right)$ can be expressed in terms of MPLs. %\claudecomment{Here we need to be more precise. What are $P$ and $s$? Are they really needed?}\paulcomment{\(s\) stands for 'spin-cover´ or 'square-root´, i.e., volume is not an iterated integral on \(\cM_{0,N+2}\) but on the finite cover \(\cM_{0,N+2}^s\) where \(\sqrt{[z_i,z_j,z_m,z_w]}\) is well-defined; \(P\) is for alternating polygon we only need \(T_{0,1,2,\ldots,n+1}\) but \(T_P\) is defined recursively}. 
More precisely, we can write 
\begin{equation}\label{eq:h_to_ALi}
    %h(\cS) = \ALi (T_{(0,\ldots, n+1)})
    h\left(\cS_z\right)=\mathrm{ALi}^{\mathcal{H}}\left(\mathrm{T}_{(0,\ldots,N+1)} (z)\right)\,,
\end{equation}
where $\mathrm{T}_{(0,\ldots,N+1)}(z)$ is defined through a purely combinatorial construction~\cite{Rudenko2020Orthoschemes}. It takes as input a point $z\in \cM_{0,N+2}$ and produces a linear combination of tensor products of `weighted letters' $[\varphi, n]$. The details of the definition of the weighted letters and the map $\mathrm{T}_{(0,\ldots,N+1)}$ are not important for the discussion here. It suffices to say that $\mathrm{T}_{(0,\ldots,N+1)}$ performs a sum over all quadrangulations of the polygon with vertices labeled $(0,\ldots,N+1)$. Each quadrangulation specifies a way of decomposing the $(N+2)$-gon into elementary quadrangles, and to every such quadrangle one associates a basic `letter' built from cross-ratios of the four vertices. A concrete implementation of this map is provided in the ancillary material of ref.~\cite{Ren:2023tuj}. 
Finally, the function $\ALi^{\mathcal{H}}$ is a linear map that assigns to \([\varphi_1,m_1] \otimes \ldots \otimes[\varphi_k,m_k]\) a combination of \emph{alternating polylogarithms}, defined by
\begin{equation}\label{eq:ALi_def}
\operatorname{ALi}_{m_1, \ldots, m_k}\!\left(\varphi_1, \ldots, \varphi_k\right) = \sum_{\epsilon_1, \ldots, \epsilon_k \in\{-1,1\}}\left(\prod_{i=1}^k \frac{\epsilon_i}{2}\right) \operatorname{Li}_{m_1, \ldots, m_k}\!\left(\epsilon_1 \sqrt{\varphi_1}, \ldots, \epsilon_k \sqrt{\varphi_k}\right)\,.
\end{equation}
The arguments $\varphi_i$ are (products of) cross ratios formed out of the $z_i$ that define a point in $\cM_{0,N+2}$, which can themselves be expressed in term of the elements of the Gram matrix via eq.~\eqref{eq:z_i_to_Q}. The cross ratios are defined by
\beq\label{eq:cr_def}
\operatorname{cr}(i_0,i_1,i_2,i_3) = \frac{(z_{i_0}-z_{i_1})(z_{i_2}-z_{i_3})}{(z_{i_0}-z_{i_3})(z_{i_1}-z_{i_2})}\,.
\eeq
Note that the definition of \(\operatorname{ALi}_{m_1, \ldots, m_k}\) only depends on the choice of the sign of the square-roots through an overall sign. It is thus important to keep a consistent choice of signs for the square-roots.

So far we have only discussed the case $D=N$ with $N$ even. Let us briefly comment on the case $D=N+1$ when \(N\) is odd. 
%Note that, using dimensional recurrence and IBP relations, all one-loop integrals can be expressed in terms of the integrals with $D=N$ and $D=N+1$. 
%The previous discussion, however, is a priori only valid for even \(N\).
One-loop integrals with $N$ even are invariant under extended dual conformal transformations. It is then easy to see that the integral with $N=D-1$ can be related to the integrals with $N=D$ where one of the points is at infinity. 
%In appendix~ \ref{app:proofs} we show %\claudecomment{Who is `One'? We need a reference, or an explicit proof!}\paulcomment{added a sketch in the appendix using Feynman parameters} 
This amounts to working with the \emph{extended Gram matrix} $Q_{\textrm{ext}}$, obtained from the Gram matrix $Q$ via 
\beq\label{eq:Qext_def}
Q_{\textrm{ext}} = \begin{pmatrix}
&&& c \\
& Q && \vdots\\
&&&c \\
c&\cdots&c &0
\end{pmatrix}\,,
\eeq
where $c$ is a non-zero constant. In the following the choice $c=\tfrac{1}{2}$ will be convenient.
When inserting this form of the Gram matrix into the general formula for the volumes of orthoschemes from ref.~\cite{Rudenko2020Orthoschemes}, individual terms may divergence (though the sum is finite).  
%\paulcomment{it's more a technicality to use Rudenko's formula (generic) and then taking the limit isn't it?}\claudecomment{Better?} 
In order to circumvent these spurious divergences, we introduce a regulator $\delta$ via, 
\beq\label{eq:Qext_def_regulated}
Q_{\textrm{ext}}^{\delta} = \begin{pmatrix}
&&& \tfrac{1}{2} \\
& Q && \vdots\\
&&&\tfrac{1}{2} \\
\tfrac{1}{2}&\cdots&\tfrac{1}{2} &\delta
\end{pmatrix}\,.
\eeq
 and taking the limit $\delta\to0$ at the end. This step is only a complication of technical nature, and we refer to ref.~\cite{Ren:2023tuj} for details. The upshot is that the volume of every orthoscheme, and thus every Feynman integral $I_N^N(Q)$ ($N$ even) or $I_N^{N+1}(Q) = I_{N+1}^{N+1} (Q_\textnormal{ext})$ ($N$ odd), can be expressed in terms of alternating MPLs, whose arguments can be expressed in terms of the variables $z_i$ in eq.~\eqref{eq:z_i_to_Q}. In particular, the ancillary file of ref.~\cite{Ren:2023tuj} contains a Mathematica code that allows one to obtain these expressions in terms of alternating MPLs in an automated way.

% !TEX root = main.tex

\section{One-loop integrals in dimensional regularisation from volumes of hyperbolic orthoschemes}
\label{sec:dimreg}

In the previous section we have seen how one can reduce the computation of one-loop Feynman integrals in integer dimensions to the computation of volumes of orthoschemes. In this section we discuss how one can use those results to obtain analytic results for the higher orders in the Laurent expansion in the dimension regulator $\eps$. We start by discussing in this section our general strategy, before we present explicit results for triangle, box and pentagon integrals in the next section.

\subsection{Auxiliary mass integration}
\label{sec:auxiliary-mass}

We consider a one-loop $N$-point integral in $D=d-2\eps$ dimensions with $N\ge 3$ and $N=d$ or $N=d-1$. We do not consider $N\le 2$, because for those cases analytic results valid to all orders in the dimensional regulator are known. We also assume that the integral is convergent. Note that for all divergent integrals in four dimensions analytic results including the higher orders in the dimensional regulator are known, so that the restriction to convergent integrals is justified. We assume that we want to compute a one-loop amplitude in the \mbox{`t~Hooft}-Veltman scheme, where the external momenta $l_i$ lie in $d$ dimensions, and only the loop momentum $l$ is taken in $D$ dimensions. We may split the loop momentum into a $d$-dimensional part $l_{\parallel}$ and a $(-2\eps)$-dimensional part $l_{\bot}$. In the `t~Hooft-Veltman scheme we have
\beq
l\cdot l_i = l_{\parallel}\cdot l_i \textrm{~~~and~~~} l_{\bot}\cdot l_i =0\,,
\eeq
and so
\beq
(l-l_i)^2 + m_i^2 = (l_{\parallel}-l_i)^2  +  m_i^2 + \eta\,,
\eeq
where we defined $\eta = l_{\bot}^2$. If we introduce spherical coordinates in the $(-2\eps)$-dimensional space, the $D$-dimensional integration measure becomes
\beq
\rd^Dl = \frac{1}{2}\, \eta^{-1-\varepsilon}\,\rd^dl_{\parallel}\,\rd\Omega_{-2\varepsilon}\,\mathrm{d}\eta\,,
\eeq
where $\rd\Omega_{-2\varepsilon}$ is the infinitesimal solid angle in $(-2\eps)$ dimensions, i.e., the integration measure on $S^{-2\eps-1}$.
The integral then becomes
\begin{equation}
\begin{split}
\label{eq:master}
    I^D_N(Q)&=\frac{1}{2}\Omega_{-2 \varepsilon}\,e^{\gamma_E\eps}\, \pi^{\varepsilon} \int_{0}^{\infty} \mathrm{d}\eta \ \eta^{-1-\varepsilon}\!\! \int \frac{\rd^{d} l_{\parallel}}{\pi^{d / 2}} \prod_{i=1}^N \frac{1}{\left(l_{\parallel}-l_i\right)^2+m_i^2+\eta}\\
    &= \frac{e^{\gamma_E\eps}}{\Gamma\!\left( -\varepsilon \right)}  \int_{0}^{\infty} \mathrm{d}\eta \ \eta^{-1-\varepsilon}\, I^{d}_N \big({Q}_{\eta} \big)\,,
\end{split}
\end{equation}
where ${Q}_{\eta} = Q_{|m_i^2\to m_i^2+\eta}$ is obtained from $Q$ by replacing $m_i^2$ by $m_i^2+\eta$ everywhere. Inserting this into eq.~\eqref{eq:Cayley}, one finds the simple relation 
\beq\label{eq:Q_eta_def}
Q_\eta = Q + \eta\,
\mathbf{1}\mathbf{1}^\top\,,
\eeq
where $Q$ is an $N\times N$ matrix, and $\mathbf{1}$ is the $N$-dimensional column vector whose entries are all 1. Equation~\eqref{eq:master} allows us to express the one-loop integral $I^D_N(Q)$ as a one-fold integral over the corresponding integral in $D=d$ dimensions, but with the squared masses of the propagators shifted by $\eta$. This implies that all propagators that appear in $I^{d}_N \big({Q}_{\eta} \big)$ are massive, and so these integrals are finite. As $N=d$ or $N=d-1$, we can use the results from ref.~\cite{Ren:2023tuj} reviewed in the previous section to obtain a representation of the integrand in terms of MPLs. Note that $I^{d}_N \big({Q}_{\eta} \big)$ is independent of the dimensional regulator $\eps$. The prefactor involving the $\Gamma$ function  vanishes as $\eps\to0$, which reflects the fact that the integral over $\eta$ develops a pole at $\eps=0$. 

The general idea is now clear: One may attempt to insert the analytic expressions for the one-loop integrals obtained in ref.~\cite{Ren:2023tuj} by computing volumes of orthoschemes, and then perform the integration over $\eta$ order by order in $\eps$ using direct integration techniques in terms of MPLs~\cite{Brown2009MasslessHigherLoop,Anastasiou:2013srw,Ablinger:2014yaa,Panzer:2014caa,hyperlogprocedures}. 
However, this naive strategy fails, for two reasons. First, the dependence on $\eta$ is hidden in the arguments of the MPLs, which involve square roots, cf. eq.~\eqref{eq:ALi_def}. There is no guarantee that direct integration techniques with MPLs will succeed in the presence of square roots. Second, we need to discuss how to extract the poles in $\eps$, because the integral over $\eta$ is divergent for very small values of $\eta$. We discuss both of these issues in turn in the remaining subsections.

%%%%%%%%%%%%%%%%%%%%%%%%%%%%%%%%%%%%%%
%%%%%%%%%%%%%%%%%%%%%%%%%%%%%%%%%%%%%%

\subsection{Dissecting  simplices into orthoschemes (2)}
\label{sec:basic_simplex}

We start by discussing how we can deal with the square roots that appear in the arguments of the alternating MPLs to make the integrand amenable to direct integration techniques. Direct integration techniques in terms of MPLs typically require one to find an ordering of the integration variables such that at every step in the integration the symbol letters are linear in the next integration variable. This criterion easily fails if the MPLs involve square roots of polynomials of the integration variables. One way to proceed can be to find a change of variables that rationalises all square roots~\cite{Besier:2018jen,Besier:2019kco}. In the present case, where we need to perform a single integral over $\eta$ in eq.~\eqref{eq:master}, we need to find a way to rationalise all the square roots involving $\eta$ in the arguments of the alternating MPLs.

If we substitute the expressions for $I^{d}_N \big({Q}_{\eta} \big)$ obtained from the formula in eq.~\eqref{eq:diss_Ren}, we find that the individual orthoschemes have a very complicated dependence on square roots involving $\eta$, and we were not able to rationalise these square roots in a way that allows us to perform the integral in eq.~\eqref{eq:master} (except possibly in the simplest cases). For this reason, we were not able to proceed with our strategy if we work with the decomposition into orthoschemes from ref.~\cite{Ren:2023tuj}. This impasse motivates the search for an alternative decomposition into orthoschemes. In what follows, we will present two distinct strategies to circumvent this issue, and we demonstrate their equivalence. 

\subsubsection{The basic simplex}
In ref.~\cite{Davydychev:1997wa} another way of dissecting the polytope associated to a one-loop Feynman integral was proposed, called the \emph{splitting of the basic simplex}.
The basic simplex encodes the kinematics of a one-loop Feynman integral~\cite{Davydychev:1997wa}. We have $\tfrac{Q_{i,j}^2}{Q_{i,i}Q_{j,j}}<1$ whenever $(m_i-m_j)^2 \le - (l_i-l_j)^2 \le (m_i+m_j)^2$.  
We can then construct a geometric representation of the masses \(m_i\) and momenta \(L_{ij} = \sqrt{-(l_i-l_j)^2}\) in an \(N\)-dimensional Euclidean space by assigning to each mass $m_i$ a vector \(\mathbf{m}_i\) of length $m_i$.
%= m_i \mathbf{a}_i\), where \(\mathbf{a}_i\) is a unit vector. 
All vectors originate from a central point called the \emph{mass meeting point}. The entire arrangement is uniquely fixed (up to rotations and reflections) by requiring that the distance between the endpoints of \(\mathbf{m}_i\) and \(\mathbf{m}_j\) is \(L_{ij}\). 
%We can arrange the masses \(m_i\) into vectors $m_i a_i$
%\claudecomment{What is $a_i$?}\paulcomment{unit vector in \(\mathbb{R}^N\); there is only one such configuration mod rotations} 
%in $N$ dimensions with unit vectors \(a_i\) meeting at one point -- the \emph{mass meeting point} -- such that the distance between the ends of \(m_i a_i\) and \(m_j a_j\) is \(\sqrt{- (l_i-l_j)^2}\). There is only one such configurations modulo rotations and mirroring.
We obtain in this way an \(N\)-simplex, called the \emph{basic simplex} in ref.~\cite{Davydychev:1997wa}. Note that the basic simplex should not be confused with the hyperbolic \((N-1)\)-simplex used to relate Feynman integrals to volumes of simplices. 

We can dissect the basic simplex into simpler quantities.
We take the mass meeting point and project it onto the opposite hyperplane, and we carry on recursively. 
%This step dissects the basic simplex (kinematics \(Q\)) into \(N!\) simplices (kinematics \(\cQ\)). Each of these simplices is by definition also a Feynman integral. Let us refer to this dissection of \(Q\) into a set by \(\operatorname{BS}(Q) \ni \cQ\). Ref.\ \cite{Davydychev:1997wa} showed that this splitting of the kinematics results in a decomposition of the Feynman integral into a sum of Feynman integrals:
This geometric dissection partitions the basic simplex, which represents the kinematics encoded by the Gram matrix \(Q\), into \(N!\) smaller simplices. Note that this Gram matrix encodes the kinematics of the Feynman integral. Each of these new simplices defines through its Gram matrix \(\cQ\) the kinematics for a corresponding Feynman integral. We denote the set of these resulting simplices by \(\operatorname{BS}(Q)\). 
%As shown in Ref.~\cite{Davydychev:1997wa}, this geometric dissection of the kinematics implies a decomposition of the Feynman integral into a sum over the new integrals:
This approach has the advantage of allowing us to identify the individual terms in the decomposition themselves as Feynman integrals of the same type, but with modified kinematics. In particular, this decomposition applies to the dimensionally regulated integrals in $D=d-2\eps$ dimensions. We can write~\cite{Davydychev:1997wa}
%If we denote by $ \operatorname{Dis}_{\infty}(Q)$ the set of simplices obtained from applying this procedure to the simplex with Gram matrix $Q$, we can write
\begin{equation}\label{eq:diss_Sym}
     I^D_N (Q) = \frac{1}{\sqrt{\det Q}} \sum_{\mathcal{Q}\in \operatorname{BS}(Q)} \operatorname{sgn}(\mathcal{Q})\,\sqrt{\det \mathcal{Q}} \, I^D_N (\mathcal{Q})\,.
\end{equation}
%This decomposition, originally derived in the physical regime, holds equally in the Euclidean regime.

It is instructive to compare eq.~\eqref{eq:diss_Sym} to eq.~\eqref{eq:diss_Ren}. We see that the main difference is that in eq.~\eqref{eq:diss_Sym} the right-hand side is a sum of Feynman integrals with different kinematics, while in eq.~\eqref{eq:diss_Ren} the quantities that appear on the right-hand side are volumes of orthoschemes, which taken individually are not necessarily Feynman integrals. Another difference between the two equations lies in the fact that eq.~\eqref{eq:diss_Sym} involves $N!$ terms, while eq.~\eqref{eq:diss_Ren} only receives contributions from $(N-1)!$ orthoschemes.\footnote{We will often refer to the quantities appearing in the right-hand side of eq.~\eqref{eq:diss_Sym} as orthoschemes, even though an interpretation as geometric objects is strictly speaking only applicable for $\eps=0$. } This increased number of terms is compensated by the other nice properties of the decomposition in eq.~\eqref{eq:diss_Sym}, which will be discussed below. %\paulcomment{might add comments that we sometimes loosely refer to the integrals on the RHS of \eqref{eq:diss_Sym} as orthoschemes even though only \(I^N_N  (\cQ)\) is usually interpreted as a geometric object}\claudecomment{Added a footnote/}

%The strategy now is that since we can calculate \(I_N^D(Q)\) from \(I_N^D(\cQ)\) with $\cQ\in\mathrm{BS}(Q)$. In fact the using the volume interpretation for \(I_N^d(\cQ)\), we see that \(\cQ\) actually corresponds to a hyperbolic orthoscheme, so that \(I_N^d(\cQ)\) computes the volume of this orthoscheme. Moreover, the orthoschemes obtained from this decomposition will have precisely the feature that we need in order to analytically evaluate the auxiliary mass integration up to $N\le 5$. Before we discuss this point, we relate our decomposition to another decomposition into orthoschemes of more geometrical nature.
%Using the auxiliary mass representation, we no longer need to further dissect the volume part of \(I_N^d(\cQ_\eta)\), and furthermore, since \(I_N^d(\cQ)\) is a Feynman integral we must also have \(\cQ_\eta = \cQ + \eta\). This will greatly simplify computations, i.e., as we will see later these integrals have a predictable number of independent square roots which can all be resolved with rationisation techniques up to \(N=5\).

\subsubsection{The projective dissection}
\label{sec:projective_dissection}
We now describe another dissection, which is more geometric in nature, but which turns out to be equivalent to the decomposition of the basic simplex from the previous subsection.
The idea is that we do not consider the \((N-1)\)-dimensional hyperbolic simplex \(\cS\) associated to the kinematics \(Q\) directly. Instead we are going to use the following trick to get a more symmetric dissection. Consider the \(N\)-dimensional hyperbolic simplex \(\cS'\) associated to \(Q_\textnormal{ext}\). This amounts to embedding the simplex \(\cS\) into \(\mathbb{H}^N\) and adding an ideal point \(P_\infty\) (ideal points correspond to zeroes on the diagonal of the Gram matrix). The simplex \(\cS'\) is then the convex hull of \(\cS\) and \(P_\infty\). We can thus use the dissection technique from section~\ref{sec:hyperbolic} using our auxiliary ideal point to start the dissection to obtain a dissection of \(\cS'\) into \(N!\) orthoschemes in \(\mathbb{H}^N\). Observe that on the locus \(\cS \hookrightarrow \mathbb{H}^{N}\) we also get a dissection of \(\cS\) into \(N!\) pieces.\footnote{In the case \(N\) odd we do not need this projection step since we consider Feynman integrals with an odd number of legs as \((N+1)\)-point functions with Gram matrix already being \(\Q_\textnormal{ext}\), giving us a natural choice for the starting vertex of the dissection.} 
%Since these pieces are faces of orthoschemes, they are themselves orthoschemes. 
In this way we obtain a novel dissection of our original simplex \(\cS\). 
This dissection, which we refer to as the \emph{projective dissection}, amounts to first taking the simplex \(\cS\) together with the projection \(P'_\infty\) of the ideal point \(P_\infty\) onto the hyperplane containing \(\cS\), and then divide \(\cS\) into \(N\) simplices \(\cS_i\) by constructing them as the convex hull of \(P'_\infty\) and each codimension-one face of \(\cS\). In a second step, we use the familiar dissection technique on each \(\cS_i\) starting from the common point \(P'_\infty\) to obtain a dissection into orthoschemes. 
We will denote the set of orthoschemes obtained from this dissection as \(\operatorname{Dis}_{P_\infty}(Q)\). This leads to the following formula for the volume and this the Feynman integral,
\begin{equation}\label{eq:diss_Infinity}
    I^N_N (Q) = \Gamma\! \left( \tfrac{N}{2} \right) \frac{1}{\sqrt{\det Q}} \sum_{\cQ\in \operatorname{Dis}_{P_\infty} \!(Q)} \operatorname{sgn}(\cQ)\operatorname{Vol}(\cQ)\,.
\end{equation}
\begin{figure}[!th]
    \centering
    \begin{tikzpicture}
        \node[inner sep=0pt] (img) {%
            \includegraphics[width=0.35\linewidth]{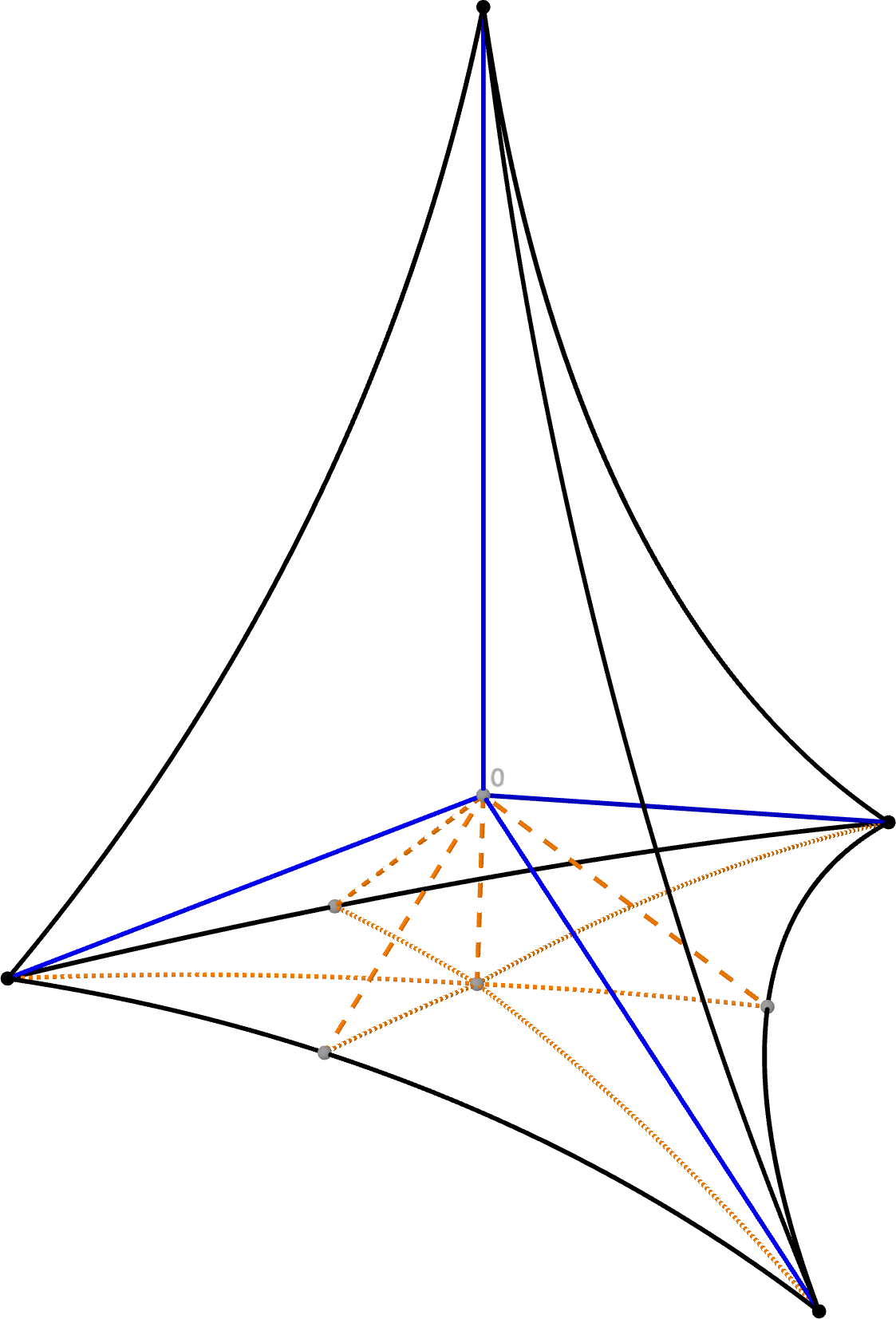}
        };
        \node[text=gray,fill=white,fill,text opacity=1,rounded corners=2pt,inner sep=1pt] at (0.55,-0.55) {$P'_\infty$};
    \end{tikzpicture}
    %\vspace{96pt}
    %\incfig[.4]{3-simplex_Diss24}
    \caption{Depiction of the projective dissection with point \(P'_\infty\) in the centre of the simplex. First the original simplex \(\mathcal{S}\) is dissected into four simplices \(\mathcal{S}_i\) (blue lines). We illustrate on the example of the simplex at the bottom how to further dissect this simplex into orthoschemes.%Dissecting a three-simplex into four other three-simplices--can be dissected into orthoschemes individually, e.g., see the lower three-simplex. \claudecomment{The previous sentence is not clear.} The special point was without loss of generality taken to be the middle of the ball for easier illustrations. %\claudecomment{there is a lot of blank space. Can this be removed?}\paulcomment{without the ball?}
    }
    \label{fig:dissection_Infty}
\end{figure}

Figure \ref{fig:dissection_Infty} illustrates this dissection for a three-simplex \(\cS\). First the three-simplex \(\cS\) is dissected into four other three-simplices \(\cS_i\) through the point \(P'_\infty\). Then, one of the \(\cS_i\) is further dissected into six orthoschemes using the dissection technique from section~\ref{sec:hyperbolic}.
%and the further dissection of one of these simplices into orthoschemes via this procedure. 
%We may then apply the decomposition into orthoschemes from ref.~\cite{Ren:2023tuj} reviewed in section~\ref{sec:setup}, with \(v_N\) being the projection of the point at infinity. 

The projective dissection has the advantage of being more symmetric than the one from ref.~\cite{Ren:2023tuj}, because it does not require one to pick one of the vertices of the original simplex to play a distinguished role. The price to pay is that we need to introduce an additional, auxiliary point. 

The projective dissection is not independent from the splitting of the basic simplex discussed in the previous subsection. In fact, the projective dissection provides a geometric interpretation of the splitting of the basic simplex at \(\varepsilon=0\). Recall that the decomposition of the basic simplex leads to a representation of the integral as a sum over \(N!\) simpler integrals, defined by the kinematics \(\mathcal{Q}\in\operatorname{BS}(Q)\). Our central claim is that the set of simplices corresponding to these kinematics is precisely the set of \(N!\) orthoschemes generated by our projective dissection, i.e., we have \(\operatorname{Dis}_{P_\infty}(Q) = \operatorname{BS} (Q)\). In appendix~\ref{eq:proj_split_equiv}  we show that the Gram matrices for the components of both decompositions are identical.

\subsubsection{Properties of the projective dissection}

Let us assume that we start from a one-loop integral whose kinematics is encoded in the Gram matrix $Q$. Adding the additional point as the projection of the point at infinity corresponds to considering the simplex (in one dimension higher) with the extended Gram matrix $Q_{\textrm{ext}}$ defined in eq.~\eqref{eq:Qext_def} (or rather its regulated version $Q_{\textrm{ext}}^{\delta}$ in eq.~\eqref{eq:Qext_def_regulated}, and then taking the limit $\delta\to0$ at the very end).
Each orthoscheme appearing in the dissection is characterised by an index sequence \(\mathbf{j}=(j_1,j_2,\ldots,j_N)\) which gives the ordering of the vertices, i.e., the first vertex is the vertex \(v_{j_1}\) of the full simplex, the second vertex is the vertex on the line inbetween \(v_{j_1}\) and \(v_{j_2}\), the third vertex is the vertex in the span of the face \(v_{j_1}\), \(v_{j_2}\), and \(v_{j_3}\), and so on. We will label the orthoschemes through \(\cQ(\mathbf{j};Q)\). We will also write $\mathcal{Q}(\mathbf{j})$ when it is clear which simplex $Q$ is being discussed, or simply $\mathcal{Q}(Q)$ when the particular orthoscheme is not important; and, when no ambiguity arises, we may even write $\mathcal{Q}$ alone.
%
%Therefore, the geometric procedure of embedding the hyperbolic simplex in a higher dimension and projecting from a point at infinity is an elegant, manifestly symmetric way to generate the exact set of kinematics that appear in the algebraic decomposition of eq.~\eqref{eq:diss_Sym}.
%Let us discuss what this dissection implies at the level of the Gram matrices. 
%
 We can then observe the several general properties, which we discuss in the remainder of this section. 
%First, as a consequence of eq.~\eqref{eq:mathcalQ_coords} and the Cauchy-Schwartz in Euclidean geometry the \(\mathcal{Q}_{ii}\) are ordered in Euclidean kinematics,
%\begin{equation}\label{eq:Euclidean_ordering}
%    0 < \mathcal{Q}_{11} < \mathcal{Q}_{22} < \ldots < \mathcal{Q}_{ii} < \mathcal{Q}_{i+1,i+1} < \ldots < \mathcal{Q}_{N,N}\,.
%\end{equation}
%\paulcomment{below belongs somewhere here}

\paragraph{Entries of the special orthoscheme matrix.}
For any orthoscheme in the projective dissection, the entries \(\cQ_{i,k}(\mathbf{j})\) with \(i \le k\) can be computed by applying the formula in ref.~\cite{Ren:2023tuj}. This yields the following formula:
\beq\label{eq:mathcalQ_coords}
    \cQ_{i,k}(\mathbf{j}) = \cQ_{k,k}(\mathbf{j}) = \frac{\det Q_{\mathbf{j}_{[k]}}}{-4 \det Q_{\mathbf{j}_{[k]}}^\textnormal{ext}}\,.
\eeq
Here \(\mathbf{j}_{[k]}=(j_1,j_2,\ldots,j_k)\) denotes the restriction of \(\mathbf{j}\) to the first \(k\) elements, \(\cQ_{I} \coloneqq \cQ_{I,I}\) where \(\cQ_{I,J}\) is the submatrix of \(\cQ\) formed by rows indexed by \(I\) and columns indexed by \(J\), and \(Q_{I}^\textnormal{ext} = (Q_I)_{\textnormal{ext}}= \begin{pmatrix}
    \cQ_I & \tfrac{1}{2}\, \mathbf{1} \\ \tfrac{1}{2}\, \mathbf{1}^\top & 0
\end{pmatrix}\). 
Observe that the last matrix element is
\beq\label{eq:mathcalQ_NN}
    \cQ_{N,N} (\mathbf{j}) = \frac{\det Q}{-4 \det Q_\textnormal{ext}}\,,
\eeq
and so it is independent of the chosen orthoscheme \(\mathbf{j}\). Furthermore, note that the Gram matrix in eq.~\eqref{eq:mathcalQ_coords} has the special form
\beq\label{eq:Gram_special}
    \cQ = 
    %\left(\cQ_{\max\{i,j\},\max\{i,j\}} \right)_{1 \le i \le N \atop 1 \le j \le N} =
    \begin{pmatrix}
        \cQ_{1,1}            & \cQ_{2,2}        & \cQ_{3,3}          & \cdots & \cQ_{N,N} \\
        \cQ_{2,2}           & \cQ_{2,2}          & \cQ_{3,3}         & \cdots & \vdots          \\
        \cQ_{3,3}          & \cQ_{3,3}        & \cQ_{3,3}          & \cdots & \vdots          \\
        \vdots           & \vdots          & \vdots           & \ddots & \vdots          \\
        \cQ_{N,N}   & \cdots          & \cdots           & \cdots & \cQ_{N,N}
    \end{pmatrix}\,.
\eeq
%It can, therefore, equally be threated as an intrinsic quantity associated to the \(N\)-simplex \(Q\).
%\claudecomment{We should meniton the special form of the Gram matrix!}

\paragraph{Additivity in the auxiliary mass parameter.}
One important property of the projective dissection is that it behaves nicely under the shifts by the auxiliary mass parameter $\eta$. We have (cf. eq.~\eqref{eq:Q_eta_def}),
\beq\label{eq:cQ+eta}
    \cQ (\mathbf{j};Q_\eta) = \cQ (\mathbf{j};Q+\eta) = \cQ(\mathbf{j};Q)+\eta\,,
\eeq
where by \(Q+\eta\) we mean that we add \(\eta\) to each entry of $Q$. 
The proof can be found in appendix \ref{app:proof_additivity}.

%\paulcomment{better notation (@paul still need to check)}
Analogously, we can also compute the sign factor which was introduced in eq.~\eqref{eq:diss_Sym}:
%Q^\textnormal{ext}_{\mathbf{j}_{[i]},\mathbf{j}_{[i-1,N+1]}}
\begin{equation}\label{eq:sgn_Factor}
    \operatorname{sgn}(\cQ (\mathbf{j}))=\sgn(\mathbf{j})\prod_{i=1}^{N}\sgn(\mathbf{j}[i]) \, \sgn \left( \frac{\det
    \begin{pmatrix}
        Q_{\mathbf{j}[i],\mathbf{j}[i-1]} & \frac{1}{2}\mathbf{1}
    \end{pmatrix}
    }{-4 \det Q^\textnormal{ext}_{\mathbf{j}_{[i]}}} \right),
\end{equation}
where \(\begin{pmatrix} Q_{\mathbf{j}[i],\mathbf{j}[i-1]} & \frac{1}{2}\mathbf{1} \end{pmatrix}\) is the matrix \(Q_{\mathbf{j}[i]}\) with the last column replace by \(\tfrac{1}{2}\mathbf{1}\).

\paragraph{Number of square roots.} Let us now discuss the square roots involving $\eta$ that appear in the expressions for the alternating polylogarithms.
First, as a consequence of eq.~\eqref{eq:mathcalQ_coords} and the Cauchy-Schwartz in Euclidean geometry, the \(\mathcal{Q}_{i,i}\) are ordered in Euclidean kinematics,
\begin{equation}\label{eq:Euclidean_ordering}
    0 < \mathcal{Q}_{1,1} < \mathcal{Q}_{2,2} < \ldots < \mathcal{Q}_{i,i} < \mathcal{Q}_{i+1,i+1} < \ldots < \mathcal{Q}_{N,N}\,.
\end{equation}
This is consistent with the ordering \(z_0<z_n<z_{n-1}<\cdots<z_1<z_{n+1}\) of the moduli space coordinates\footnote{Since we are using the equivalent conventions of ref.~\cite{Ren:2023tuj} this differs to what is written in ref.~\cite{Rudenko2020Orthoschemes}} of the hyperbolic locus \(\mathcal{M}_{0, m+2}^h \subset \mathcal{M}_{0, m+2}\) \cite[cf.\ Def.\ 6.11]{Rudenko2020Orthoschemes}, because the coordinates \(z_i\) simplify to
\beq\label{eq:special_z}
    z_i = \frac{\cQ_{i,N}^2}{\cQ_{i,i},\cQ_{N,N}} = \frac{\cQ_{N,N}}{\cQ_{i,i}}.
\eeq
Second, as a consequence of the structure of the map $\mathrm{T}_{(0,\ldots,N+1)}$, only the cross-ratios \(\operatorname{cr}(i_0,i_1,i_2,i_3)\) in eq.~\eqref{eq:cr_def} with \(i_k-i_{k+1}\) being odd (and products thereof) enter the arguments of the alternating polylogarithms. Using eq.~\eqref{eq:special_z} it is easy to show that 
\begin{equation}\bsp
    \operatorname{cr}(0,i_1,i_2,N+1) &= \frac{\cQ_{i_2,i_2}}{\cQ_{i_1,i_1}-\cQ_{i_2,i_2}}\,,\\
    \operatorname{cr}(i_0,i_1,i_2,N+1) &=\frac{\cQ_{i_2,i_2} (\cQ_{i_0,i_0}-\cQ_{i_1,i_1})}{\cQ_{i_0,i_0} (\cQ_{i_1,i_1}-\cQ_{i_2,i_2})}\,,\\
    \operatorname{cr}(0,i_1,i_2,i_3) &=\frac{\cQ_{i_2,i_2}-\cQ_{i_3,i_3}}{\cQ_{i_1,i_1}-\cQ_{i_2,i_2}}\,,\\
    \operatorname{cr}(i_0,i_1,i_2,i_3) &=\frac{(\cQ_{i_0,i_0}-\cQ_{i_1,i_1}) (\cQ_{i_2,i_2}-\cQ_{i_3,i_3})}{(\cQ_{i_0,i_0}-\cQ_{i_3,i_3}) (\cQ_{i_1,i_1}-\cQ_{i_2,i_2})}\,.
\esp\end{equation}
Since \(i_k-i_{k+1}\) is odd, \(i_m\) is even for \(m\) even and vice versa. When we now consider external kinematics to be \(Q_\eta\), we observe that only the following cross-ratios depend on \(\eta\): 
\beq
    \operatorname{cr}(0,i_1,i_2,N+1)_{|Q\mapsto Q_\eta} = \frac{\cQ^\eta_{i_2,i_2}}{\cQ^\eta_{i_1,i_1}-\cQ^\eta_{i_2,i_2}}=\frac{\cQ_{i_2,i_2}+\eta}{\cQ_{i_1,i_1}-\cQ_{i_2,i_2}}\,,
\eeq
and
\beq\bsp
    \operatorname{cr}(i_0,i_1,i_2,N+1)_{|Q\mapsto Q_\eta} &=\frac{\cQ^\eta_{i_2,i_2} (\cQ^\eta_{i_0,i_0}-\cQ^\eta_{i_1,i_1})}{\cQ^\eta_{i_0,i_0} (\cQ^\eta_{i_1,i_1}-\cQ^\eta_{i_2,i_2})}\\
    &=\frac{(\cQ_{i_2,i_2} +\eta)(\cQ_{i_0,i_0}-\cQ_{i_1,i_1})}{(\cQ_{i_0,i_0}+\eta) (\cQ_{i_1,i_1}-\cQ_{i_2,i_2})}\,,
\esp\eeq   
where we defined \(\cQ^\eta_{i,j}\coloneqq (\cQ_{\eta})_{i,j}\). The arguments of the alternating polylogarithms are square roots of products of cross ratios, and we see that only the square-roots $ 
    \sqrt{\cQ_{i_2,i_2}+\eta}$ and $\sqrt{\tfrac{\cQ_{i_2,i_2}+\eta}{\cQ_{i_0,i_0}+\eta}}$ appear inside the arguments.
Both \(i_0\) and \(i_2\) are even integers, and so at most \(\tfrac{N}{2}\) distinct square-roots can appear in the calculation of higher orders in \(\varepsilon\)  of an \(N\)-gon for \(N\) even. Note that this statement is independent of the order in the $\eps$-expansion.  The above derivation is also applicable to the case \(N\) odd, because these simplices can be seen as limits of \((N+1)\)-gon integrals. In particular, the only difference is that the moduli space coordinates \(z_i\) now become
\beq
    z_i = \frac{\cQ_{i,N+1}^2}{\cQ_{i,i} \delta} = \frac{1}{4 \cQ_{i,i} \delta} \,,
\eeq
according to eq.~\eqref{eq:Qext_def_regulated}.
In the end we then have to take the limit \(\delta \downarrow 0\). The results vary only in minor distinctions, such that the key takeaway stays the same, i.e., only even integers \(i_0\) and \(i_2\) may contribute to the square roots involving \(\eta\). In total we deduce that for general \(N\) at most  \(\lfloor \tfrac{N}{2} \rfloor\) distinct square roots can appear in the expression \(I_N^d(Q_\eta)\).
%On the moduli space coordinates this limit corresponds \paulcomment{@paul rework this this is too handwavy we haven't explained the \(cQ_{N+1,N+1} \uparrow \infty\) limit} to identifying \(z_{N+1}\) with \(z_0\) (this is a consequence of taking the limit \(\cQ_{N+1,N+1} \to \infty\)), which can be shown to reduce the Feynman parameter representation of the \((N+1)\)-gon integral to the desired \(N\)-gon integral for our special configuration of kinematics. 
%This means fewer cross-ratios can contribute, and our counting argument still persists, i.e., for general \(N\) at most  \(\lfloor \tfrac{N}{2} \rfloor\) distinct square roots appear in the expression \(I_N^d(Q_\eta)\).

We can always simultaneously rationalise two square roots that are linear in $\eta$. Hence, we can always rationalise all the $\eta$-dependent square roots, whenever \(\lfloor \tfrac{N}{2} \rfloor \le 2\), which happens for $N\le 5$. In other words, we expect that we can always evaluate the integral over $\eta$ using direct integration techniques for integrals with $N\le 5$. We have checked that the hexagon integral ($N=6$) saturates the upper bound $\lfloor \tfrac{N}{2}\rfloor = 3$, and so it cannot be computed using our method. 
We will explicitly perform computations for triangle, box, and pentagon integrals in section~\ref{sec:examples}. Before that, however, we need to discuss how to perform the expansion in $\eps$ of the integral in eq.~\eqref{eq:master}.

\subsection{Pure functions and the Laurent expansion of eq.~\eqref{eq:master}}

It is expected that one-loop integrals with $N=d$ or $N=d-1$ can be cast in the form
\begin{equation}\label{eq:I_pure}
    I^{D}_N(Q) = \left(\frac{-\iu}{2}\right)^{\tfrac{d}{2}+1}\, \frac{J^{D}_N(Q)}{\sqrt{\left|\det Q^{(N)}\right|}}\,,
\end{equation}
%\paulcomment{I would add an additional factor here: \(  I^{D}_N(Q) = \left(\frac{-\iu}{2}\right)^{\tfrac{d}{2}+1}\,\frac{J^{D}_N(Q)}{\sqrt{\det Q^N}} \) to get expressions with integer factors in front of G's}\claudecomment{OK}
where we defined 
\beq
Q^{(N)} = \left\{\begin{array}{ll}
Q\,,&\textrm{ if $N$ is even}\,,\\
Q_{\textrm{ext}}\,,&\textrm{ if $N$ is odd}\,.
\end{array}\right.
\eeq
The function $J^{D}_N(Q)$ admits a Laurent expansion,
\beq
J^{D}_N(Q) = \sum_{k\ge 0}\eps^k\,J^{d}_{N,k}(Q)\,,
\eeq
and the Laurent coefficients $J^{d}_{N,k}(Q)$ are expected to be pure functions in the sense of ref.~\cite{Arkani-Hamed:2010pyv}, i.e., they are expected to be $\mathbb{Q}$-linear combinations of MPLs of weight $\tfrac{d}{2}+k$. We now discuss how we can recast eq.~\eqref{eq:master} into a form that gives an integral representation directly for the pure functions $J^{d}_{N,k}(Q)$. 

We start by letting $\eps\to0$ in eq.~\eqref{eq:I_pure} to obtain the function that enters the integrand in eq.~\eqref{eq:master}. We obtain
\begin{equation}
    I^{d}_N(Q_{\eta}) = \left(\frac{-\iu}{2}\right)^{\tfrac{d}{2}+1}\,\frac{J^{d}_{k,0}(Q_{\eta})}{\sqrt{\big|\det Q^{(N)}_\eta \big|}}\,,
\end{equation}
where we defined $Q^{(N)}_\eta = (Q^{(N)})_{|m_i^2\to m_i^2+\eta}$. We also define
\beq
f_N(\eta,Q) = \frac{\det Q^{(N)}_\eta}{\det Q^{(N)}}\,.
\eeq
The function \(f_N(\eta,Q)\) is determined by the structure of \(\det Q(\eta)\), which -- using an intermediate result established in appendix \ref{app:proof_additivity} -- can be simplified to:
\begin{equation}
    \det Q^{(N)}_\eta= \begin{cases}
        \det Q - 4 \det Q_\textnormal{ext}\, \eta \,,& N \textnormal{ even}\,,\\
        \det Q_\textnormal{ext}  \,, & N \textnormal{ odd}\,.
    \end{cases}
\end{equation}
This gives the following very simple expression,
\beq\label{eq:fNQ}
f_N (\eta,Q) =  \begin{cases}
1- \frac{4 \det Q_\textnormal{ext}}{\det Q} \, \eta\,, & N \textrm{ even}\,,\\
1\,,&\, N \textrm{ odd}\,.
\end{cases}
\eeq
%\claudecomment{I think we should use $f_N(\eta,Q)$ consistently everywhere.}
We can then write:
\beq
    J^{D}_{N} (Q) = \frac{e^{\gamma_E\eps}}{\Gamma (-\varepsilon) } \int_0^\infty \frac{\mathrm{d}\eta}{\sqrt{f_N(\eta, Q)}}\,{\eta}^{-1-\varepsilon}\,J^{d}_{N,0}\bigl(Q_{\eta}\bigr)\,.
\eeq

At this point we could proceed in two different ways, depending on whether we use eq.~\eqref{eq:diss_Sym} or eq.~\eqref{eq:diss_Infinity}.
We can check that these two approaches lead to equivalent results. If we apply the projective  dissection from section~\ref{sec:projective_dissection}, we obtain
\beq\bsp\label{eq:JdN0_proj_diss}
    J^{d}_{N,0}(Q_\eta)&\, = \sum_{\cQ(Q_\eta)\in \operatorname{Dis}_{P_\infty} \!(Q_\eta)} \operatorname{sgn}(\cQ(Q_\eta))J^{d}_{N,0}(\cQ(Q_\eta))\\ &\,= \sum_{\cQ\in \operatorname{Dis}_{P_\infty} \!(Q)} \operatorname{sgn}(\cQ)J^{d}_{N,0}(\cQ_\eta)\,,
\esp\eeq
where we used that eqs.~\eqref{eq:mathcalQ_NN} and~\eqref{eq:cQ+eta} imply that \(\operatorname{sgn}(\cQ (Q_\eta) = \operatorname{sgn}(\cQ (Q) ) \) is independent of \(\eta\) and \(\cQ(Q_\eta) = (\cQ (Q) )_\eta = \cQ_\eta\). 
We can insert eq.~\eqref{eq:JdN0_proj_diss} into the auxiliary mass integration in eq.~\eqref{eq:master} to obtain the following expression for the dimensionally-regularised integral:
\beq\label{eq:dissection_proj_aux}
    J^{D}_{N} (Q) = \frac{e^{\gamma_E\eps}}{\Gamma (-\varepsilon) }  \int_0^\infty \frac{\mathrm{d}\eta}{\sqrt{f_N(\eta, Q)}}\,{\eta}^{-1-\varepsilon}\, \sum_{\cQ\in \operatorname{Dis}_{P_\infty} \!(Q)} \operatorname{sgn}(\cQ) J^{d}_{N,0}(\cQ_\eta) \,.
\eeq
If instead we use the dissection from eq.~\eqref{eq:diss_Sym} based on splitting the basic simplex, we obtain
\beq\label{eq:splitting_basic_aux}
    J_N^D (Q) = \sum_{\cQ \in \operatorname{BS (Q)}} \operatorname{sgn} (\cQ) J_N^D (\cQ)\,.
\eeq 
We can express \( J_N^D (\cQ)\) via the representation involving the integration over the auxiliary mass:
\beq
    J^{D}_{N} (Q) =  \sum_{\cQ \in \operatorname{BS (Q)}} \sgn \cQ \, \frac{ e^{\gamma_E \eps} }{\Gamma (-\varepsilon) } \int_0^\infty \frac{\mathrm{d}\eta}{\sqrt{f_N(\eta;\cQ)}}\,{\eta}^{-1-\varepsilon}\,J^{d}_{N,0}(\cQ_\eta).
\eeq
To compare eqs.~\eqref{eq:dissection_proj_aux} and~\eqref{eq:splitting_basic_aux} we need to compute \(f_N(\eta;\cQ)\). First we can compute the determinant of the orthoscheme \(\cQ^{(N)}_\eta\):
\beq
    \det \cQ^{(N)}_\eta= \begin{cases}
      (\cQ_{N,N}+\eta) \prod_{i=1}^{N-1}  (\cQ_{i,i}-\cQ_{i+1,i+1})\,, & N \textnormal{ even}\,,\\
        -\dfrac{1}{4}\,\prod_{i=1}^{N-1}  (\cQ_{i,i}-\cQ_{i+1,i+1})\,, & N \textnormal{ odd}\,,
    \end{cases}
\eeq
such that 
\beq\label{eq:fNcQ}
f_N (\eta,\cQ) =  \begin{cases}
1 + \frac{\eta}{\cQ_{N,N}}\,, & N \textrm{ even}\,,\\
1\,,&\, N \textrm{ odd}\,,
\end{cases}
\eeq
which is the same for all \(\cQ\in\operatorname{BS (Q)}\), because \(\cQ_{N,N}\) is independent of the chosen orthoscheme. We conclude that, because of eq.~\eqref{eq:mathcalQ_NN}, the expressions in eqs.~\eqref{eq:fNQ} and \eqref{eq:fNcQ} are equal, i.e.,
\beq
    f_N(\eta; Q) = f_N (\eta; \cQ) \quad \forall \cQ \in \operatorname{BS (Q)}\,,
\eeq
such that also eqs.~\eqref{eq:dissection_proj_aux} and \eqref{eq:splitting_basic_aux} are equal upon inspection, since we have also shown \(\operatorname{Dis}_{P_\infty} \!(Q) = \operatorname{BS (Q)}\) in appendix \ref{eq:proj_split_equiv}.

The integration variable $\eta$ is dimensionful. We render it dimensionless
by the change of variables \(\eta= \mu\,\eta'\) with \(\mu>0\). A natural choice for \(\mu>0\) would be \(\cQ_{N,N}\), because it is the same regardless of which orthoscheme in the dissection we consider. We prefer not to use this choice here, because for the triangle and pentagon integrals other choices turn out to be more convenient. 
Putting everything together and defining $\mathscr{J}_N(Q)  = \mu^\eps\,e^{-\gamma_E\eps} \, \Gamma(\eps)\, J^{D}_{N} (Q)$, we find
\begin{equation}\label{eq:curlyJ}
    \mathscr{J}_N(Q)=\int_0^\infty \frac{\mathrm{d}\eta'}{\sqrt{g_N(\eta')}}\,{\eta'}^{-1-\varepsilon}\,J^{d}_{N,0}\bigl(Q_{\eta'}\bigr),
\end{equation}
where we defined \(g_N(\eta') \coloneqq f_N (\mu \eta',Q)\) and \(Q_{\eta'}\) denotes \(Q_{\mu\eta'}\) with slight abuse of notation. From now on we drop the prime and write \(\eta\) for the (dimensionless) integration variable obtained after \(\eta=\mu\eta'\).
%; the original dimensionful variable will not appear further if not indicated otherwise. 
%Note that the Laurent coefficients of \(\mathscr{J}_N(Q)\) are pure functions.

The integral in eq.~\eqref{eq:curlyJ} develops a singularity as $\eps\to0$, and so we cannot perform a naive Taylor expansion in $\eps$ in the integrand. We now discuss how we can obtain the Laurent expansion in $\eps$. Naively, the differential form $\tfrac{\rd\eta}{\eta}$ has a simple pole at $\eta=0$ and $\eta=\infty$. The limit $\eta\to\infty$ corresponds to the large-mass limit of the Feynman integral, and one can check that the integral $I_N^d(Q_\eta)$ in the integrand vanishes as $\eta\to\infty$. Hence, there is no pole at the integration boundary $\eta=\infty$. We will see this explicitly when we discuss the examples with $N\le 5$ in section~\ref{sec:examples}. We thus only need to discuss the divergence at the integration boundary $\eta=0$. To deal with this divergence, we use the distributional expansion,
\beq
\eta^{-1-\eps} = -\frac{1}{\eps}\,\delta(\eta) + \sum_{k=0}^{\infty}\eps^k\,\left[\frac{\log^k\eta}{\eta}\right]_+\,,
\eeq
where we introduced the plus-distributions acting on test functions $\varphi(\eta)$ by
\beq
\int_0^1\rd \eta\,\varphi(\eta)\,\left[\frac{\log^k\eta}{\eta}\right]_+ := \int_0^1\rd\eta\,\frac{[\varphi(\eta)-\varphi(0)]\,\log^k\eta}{\eta}\,.
\eeq
This gives,
\beq\bsp\label{eq:J_N_regulated}
 \mathscr{J}_N(Q)=-\frac{1}{\eps}\,J_{N,0}^d(Q) +\sum_{k=0}^{\infty}\eps^k&\left\{\int_0^1 \frac{\mathrm{d}\eta}{\eta} \,\log^k\eta\,\left[\frac{1}{\sqrt{g_N(\eta)}}\,  J^{d}_{N,0} (Q_\eta)-J_{N,0}^d(Q)\right]\right.\\
 +&\,\left.\int_1^{\infty} \frac{\mathrm{d}\eta}{\eta\,\sqrt{g_N(\eta)}} \,\log^k\eta\,J^{d}_{N,0} (Q_\eta)\right\}\,.
 \esp\eeq
The remaining integrals are absolutely convergent. Since we know from the previous subsection that we can rationalise all the square roots for $N\le 5$, we conclude that these integrals are amenable to direct integration techniques in terms of MPLs. It turns out to be more convenient to perform the regularisation and the distributional expansion after the change of variables that rationalises the square roots. We will discuss this on the concrete examples in section~\ref{sec:examples}. We stress that the distinction between different $\eps$ orders lies in the exponent of $\log\eta$, while the remaining terms in the integrand are the same, independently of the order in the $\eps$ expansion. It follows that no new challenges will arise for higher orders in $\eps$, and direct integration techniques will always succeed, independently of the order in the $\eps$-expansion.

\section{Results for triangle, box and pentagon integrals}
\label{sec:examples}
In the previous section we have argued that we can cast one-loop integrals in dimensional regularisation in the form of an integral over the auxiliary mass parameter $\eta$, and that for $N\le 5$, we can perform all integrations using direct integration techniques for MPLs order by order in $\eps$. In this section we apply this idea to obtain explicit results for triangle and box integrals in $D=4-2\eps$ dimensions and pentagon integrals in $D=6-2\eps$ dimensions. In all cases we focus on the fully generic case, where all external legs and all propagators are massive, and all masses are different. These are typically the most complicated cases, because they depend on the largest number of scales. All other configurations, where some of the masses are equal or zero, can be obtained using a similar approach, or by taking appropriate limits. %In appendix~\ref{ap:massless} we illustrate on an example how to take these limits.

\subsection{Results for triangle integrals}
\label{sec:triangles}

Let us start by discussing the one-loop triangle integral. While many results for one-loop triangle integrals, including higher orders in the dimensional regulator, have been obtained in the literature, we include this computation for completeness. We discuss in detail the triangle depending on six external scales, namely three distinct external masses and three distinct propagator masses. Using eq.~\eqref{eq:master}, we see that this integral can be cast in the form of a sum of six orthoschemes \(\cS_{\cQ_{\textnormal{ext}}}\), each described by a Gram matrix of the form
\beq\label{eq:triangle_Gram}
\cQ = \begin{pmatrix}
\cQ_{1,1} & \cQ_{2,2} & \cQ_{3,3}\\
\cQ_{2,2} & \cQ_{2,2} & \cQ_{3,3}\\
\cQ_{3,3} & \cQ_{3,3} & \cQ_{3,3}
\end{pmatrix}\,.
\eeq
We thus see that each orthoscheme is a function of only the three variables $(\cQ_{1,1}, \cQ_{2,2},\cQ_{3,3})$. From eq.~\eqref{eq:Cayley} we know that each $\cQ_{i,i}$ has the dimension of a mass squared, and so the only non-trivial functional dependence can be through two dimensionless ratios. It will be convenient to define the quantities,
\begin{align}\label{eq:xy_Tri}
    x= \sqrt{\frac{\cQ_{2,2}-\cQ_{1,1}}{\cQ_{2,2}}} \textnormal{~~~and~~~}  y=\iu\ \sqrt{\frac{\cQ_{3,3}-\cQ_{2,2}}{\cQ_{2,2}}}\,,
\end{align}
which are the natural arguments of the volume
\beq\label{eq:tri_eps=0}\bsp
    J_{3,0}^4 (\cQ) &= 4 \perR \ALi_{1,1} \left(\tfrac{1}{x^2},y^2 \right)\\
    &= A_{xy} \left\{ [G(1; x) - G(1; y)] G(y; x) - G(1, y; x) \right\} \,,
\esp\eeq
where we defined the anti-symmetrisers
\beq\label{eq:A_xy_def}
A_x f(x) = f(x)-f(-x) \textrm{~~~and~~~}  A_{x y} = A_x A_y = A_yA_x\,.  
\eeq
To simplify the notations we sometimes omit the anti-symmetrizers and indicate the absence with a hat, i.e., \(K = A_{xy\ldots}\hat{K}\), if it is clear from the context over which variables \(xy\ldots\) is anti-symmetrized. Note that \(\hat{K}\) is not unique. For example, instead of \eqref{eq:tri_eps=0} we might just write:
\beq
     \hat{J}_{3,0}^4 (\cQ) = [G(1; x) - G(1; y)] G(y; x) - G(1, y; x)\,.
\eeq

In the auxiliary mass integration we instead need \(J_{3,0}^4 (\cQ_\eta) = 4 \perR \ALi_{1,1} \left(\tfrac{1}{x_\eta^2},y_\eta^2 \right)\) with adjusted arguments 
\beq\bsp
x_\eta &\,\coloneqq x|_{\cQ_{i,i}\to \cQ_{i,i}+ \mu \eta} = \frac{x}{\sqrt{1+\tfrac{\mu}{\cQ_{2,2}} \eta }}\,,\\
y_\eta &\,\coloneqq y|_{\cQ_{i,i}\to \cQ_{i,i}+\mu\eta} = \frac{y}{\sqrt{1+\tfrac{\mu}{\cQ_{2,2}}\eta}}\,.
\esp\eeq
For simplicity we choose \(\mu = \cQ_{2,2}\), such that the square-root becomes \(\sqrt{1+\eta}\).
%For simplicity we change coordinates from \(\eta \mapsto \eta^\prime = \tfrac{\eta}{\cQ_{2,2}}\) such that our integration goes over a dimensionless variable. With slight abuse of notation we might write the transformed \(x_{\eta^\prime} = x_{\cQ_{2,2} \eta}\) and \(y_{\eta^\prime} = y_{\cQ_{2,2} \eta}\) as \(x_{\eta^\prime}\) and \(y_{\eta^\prime}\) by setting the new variable as the index -- we will do this also in the following for the remaining changes of variables.
%\paulcomment{Triangle at \(\varepsilon = 0\) is given by \(\perR \ALi_{1,1} (1/x^2,y^2)\) with \(x,y\) as above. We get \(I(\eta)\) from this by replacing \(\cQ\) with \(\cQ_\eta = \cQ + \eta\) which results in \(x_\eta = x/\sqrt{1+\eta/\cQ_{22}} \) and \(y\) analogously. First change of variables \(\eta \mapsto \cQ_{2,2} \eta^\prime\) giving an overall factor \(\cQ_{2,2}^{-\varepsilon}\). Second change of variables \(\eta^\prime \mapsto u/(1-u)\) ...}
%\claudecomment{The next sentence is unchanged, because I do not understand it. How cna I shift $\cQ_{22}$, which is a constant?!}\paulcomment{Bad notation from my side: Basically in aux-mass integration we need \(I(\eta)\). From \(I(0) = f(\cQ_{11},\ldots)\) we can also get \(I(\eta) = f(\cQ_{11}+\eta,\ldots)\); hence, by shift I mean take \(\eta = 0\) result and make it \(I(\eta)\)}
%we observe that under the shift (and factoring out \(\Q_{2,2}\))\paulcomment{need to make clear that we have \(\cQ_{2,2}^{-\varepsilon}\) in the measure now} \(\Q_{2,2}\to\Q_{2,2}(1+\eta)\) followed by \(\eta\mapsto u/(1-u)\):
We let \(\eta = u/(1-u)\), and we obtain
\beq\bsp
    x_{\eta} &\,= \frac{x}{\sqrt{1+\eta}}= x \ \sqrt{1-u}\,,\\
    y_{\eta} &\,= \frac{y}{\sqrt{1+\eta}} =y \ \sqrt{1-u}\,.
\esp\eeq
%Observe that under this transformation the last term in \eqref{eq:TriEucl} does not have any square-roots; therefore, we would only need to rationalize the first part -- however, we will also use our found parametrization on the complete expression since it does not yield a more consice due to cross-cancellations. 
The integration region changes from $\eta\in[0,\infty]$ to $u\in[0,1]$. The square root $\sqrt{1-u}$ can be rationalised via the change of variables $u=1 - (1-t)^2$, and the new variable $t$ is still integrated over the range $[0,1]$. This change of variables is sufficient to rationalise all square roots in the integrand, to all orders in $\eps$. We also have 
\begin{align}
    {\Df{\eta}}\ \eta^{-1-\varepsilon} %&\xrightarrow{u} \Df u\ \left( \frac{1}{u} + \frac{1}{1-u} \right) \left( \frac{1-u}{u} \right)^\varepsilon\\
     &=\Df t\ \left( \frac{1}{t} - \frac{1}{2-t} +\frac{2}{1-t} \right) \left[ \frac{(1-t)^2}{t (2-t)} \right]^\varepsilon  \,.\label{eq:Tri_t-trafo}
\end{align}
Note that we get a spurious divergence at $t=2$ (which is the second preimage of $u=0$ under the 2-to-1 map from $t$ to $u$), which lies outside of the integration region \([0,1]\). The pole at \(t=1\) corresponds to the singularity at \(\eta = \infty\). In the previous section we have argued that this singularity cancels when the vanishing of \(J^{4}_{3,0}(\cQ_\eta)\) is taken into account. 
We therefore only need to deal with the pole at $t=0$.
In the following we denote by $\cQ_t$ the quantity obtained from $\cQ_{\eta}$ after inserting the change of variables from $\eta$ to $t$. The expression in eq.~\eqref{eq:curlyJ} can then be cast in the form
\begin{equation}
    \begin{split}
\label{eq:tri_int}        \mathscr{J}_3(\cQ) &= \int_0^1 \Df t\ \left( \frac{1}{t} - \frac{1}{2-t} +\frac{2}{1-t} \right) \left( \frac{(1-t)^2}{t (2-t)} \right)^\varepsilon J_{3,0}^4(\cQ_t)\\
        &=\int_0^1 \Df t\ t^{-1-\varepsilon} J_0(t,\eps) + \int_0^1 \Df t\ \left(\frac{2}{1-t} - \frac{1}{2-t} \right) J_\infty(t,\eps) \,,
    \end{split}
\end{equation}
where we defined
\beq\bsp\label{eq:J0_Ji_def}
 J_0(t,\eps) &\,= \left[ \frac{(1-t)^2}{(2-t)} \right]^\varepsilon J(t) = \left[ \frac{(1-t)^2}{(2-t)} \right]^\varepsilon J_{3,0}^4(\cQ_t)\,,\\
J_\infty(t,\eps)&\, = \left[ \frac{(1-t)^2}{t (2-t)} \right]^\varepsilon J(t) = \left[ \frac{(1-t)^2}{t (2-t)} \right]^\varepsilon J_{3,0}^4(\cQ_t)\,.
\esp\eeq
The functions $J(t)$ and $J(1-t)$ can be obtained from the triangle integral in exactly four dimensions and then passing to a fibration basis for the MPLs where the integration variable $t$ appears as the last integration variable. The functions $\hat{J}(t)$ and $\hat{J}(1-t)$ can then be written as:
\beq\bsp\label{eq:triangle_fiber}
    \hat{J}(t) % &\,= J_{3,0}^4(\cQ) + \widetilde{J}(t)\\
    &\,= \hat{J}_{3,0}^4(\cQ) + G\big(1 + \tfrac{1}{y}, 1 - \tfrac{1}{x}; t\big) + G(1; x ) \, G\big(1 + \tfrac{1}{y}; t\big)\,, \\
    \hat{J}(1 - t) &\,=  G\big(\tfrac{1}{y}, -\tfrac{1}{x}; t\big)\,.%+ i \pi A_y G\left( \frac{1}{y}; t\right)
\esp\eeq

We can now rearrange terms in eq.~\eqref{eq:tri_int} to arrive at an expression that can be expanded in $\eps$ under the integral sign using the distributional expansion of \(t^{-1-\eps}\),
\begin{equation}\label{eq:TriInt}
    \begin{split}
        \mathscr{J}_3(\cQ) 
        &= -\frac{1}{\varepsilon}\  2^{-\varepsilon} J_{3,0}^4(\cQ) + \int_0^1 \Df t \ t^{-1-\varepsilon} \Big[ J_0 (t,\eps) - 2^{-\varepsilon} J_{3,0}^4(\cQ) \Big]\\
        &\,\qquad + \int_0^1 \Df t\, \left[\frac{2}{1-t} - \frac{1}{2-t} \right] J_\infty (t,\eps)\\
        &=  -\frac{1}{\varepsilon}\  2^{-\varepsilon} J_{3,0}^4(\cQ) + \int_0^1 \frac{\Df t}{t} \left[ t^{-\varepsilon} J_0 (t,\eps) - (2t)^{-\varepsilon} J_{3,0}^4(\cQ) + 2J_\infty (1-t,\eps)\right]\\
        &\,\qquad - \int_0^1 \frac{\Df t}{2-t}  J_\infty (t,\eps)\,.
    \end{split}
\end{equation}
All the integrals appearing in the last line of eq.~\eqref{eq:TriInt} are convergent and can be expanded in $\eps$ under the integral sign. 
We stress that eq.~\eqref{eq:TriInt} is valid to all orders in $\eps$. %$J_{3,0}^4(\cQ)$ is the value of the triangle integral in exactly $D=4$ dimensions, which is well known in the literature. The functions $J(t)$ and $J(1-t)$ in eq.~\eqref{eq:J0_Ji_def} can be obtained from the triangle in exactly 4 dimensions.

%The functions $J(t)$ and $J(1-t)$ in eq.~\eqref{eq:J0_Ji_def} can be obtained from the triangle in exactly 4 dimensions, i.e., $\hat{J}(t)$ and $\hat{J}(1-t)$ can be expressed in the following from:
%\beq\bsp\label{eq:triangle_fiber}
%    \hat{J}(t) % &\,= J_{3,0}^4(\cQ) + \widetilde{J}(t)\\
%    &\,= \hat{J}_{3,0}^4(\cQ) + G\big(1 + \tfrac{1}{y}, 1 - \tfrac{1}{x}; t\big) + G(1; x ) \, G\big(1 + \tfrac{1}{y}; t\big)\,, \\
%    \hat{J}(1 - t) &\,=  G\big(\tfrac{1}{y}, -\tfrac{1}{x}; t\big)\,.%+ i \pi A_y G\left( \frac{1}{y}; t\right)
%\esp\eeq
%
%We note that the expressions in eq.~\eqref{eq:triangle_fiber} are obtained starting from the result in terms of MPLs for the triangle integral $ J_{3,0}^4(\cQ_t)$ in exactly $D=4$ dimensions, and then passing to a fibration basis for the MPLs where the integration variable $t$ appears as the last integration variable. 
It is now an easy exercise to expand eq.~\eqref{eq:TriInt} to any desired order in $\eps$ and to perform all integrations in terms of the definition of MPLs in eq.~\eqref{eq:MPL_def}. For example, if we write
\beq\label{Jhat30}
\mathscr{J}_3(\cQ) = -\frac{1}{\eps}\,J_{3,0}^4(\cQ)+\sum_{k=0}^\infty\eps^k\,\mathscr{J}_{3,k}(\cQ) \,,
\eeq
we find for the coefficient of $\eps^0$,
%\beq\label{eq:Jred_Tri}
%    \begin{split}
%        \hat{\mathscr{J}}_{3,0}(\cQ) &= G\big(0,1+\tfrac{1}{y},1-\tfrac{1}{x};1\big)+2\,G\big(0,\tfrac{1}{y},-\tfrac{1}{x};1\big) + G( 1;x) \,G\big(0,1+\tfrac{1}{y};1 \big)\\%+2 \iu \pi A_y G\left(0, \frac{1}{y};1\right)
%        &\qquad + G\big(2,1+\tfrac{1}{y},1-\tfrac{1}{x};1\big) + G( 1;x) \, G\big(2,1+\tfrac{1}{y};1 \big)\\
%        &\equiv  -(G(y, x) G(1, -1, y) ) - G(y, x) G (1, 1, y) \\
%        &\qquad+ G(y, 1, -1, x) + G(y, 1, 1, x)  \qquad \mod \ker A_{xy}\,,
%    \end{split}
%\end{equation}
\begin{equation}\label{eq:Jred_Tri}
\begin{split}
\hat{\mathscr{J}}_{3,0}(\cQ) = %G\big(0,1+\tfrac{1}{y},1-\tfrac{1}{x};1\big)+2\,G\big(0,\tfrac{1}{y},-\tfrac{1}{x};1\big)
%+ G( 1;x) \,G\big(0,1+\tfrac{1}{y};1 \big)\\
%&\qquad + G\big(2,1+\tfrac{1}{y},1-\tfrac{1}{x};1\big) + G( 1;x) \, %G\big(2,1+\tfrac{1}{y};1 \big)\\
%&\equiv  
%%%-G(y; x)\,G(1, -1; y)  - G(y; x)\,G(1, 1; y) \\
%%%&\qquad + G(y, 1, -1; x) + G(y, 1, 1; x) 
-G(y; x)\,( G(1, -1; y) + G(1, 1; y)) + G(y, 1, -1; x) + G(y, 1, 1; x) 
%\pmod{\ker A_{xy}}
\,.
\end{split}
\end{equation}
%where we brought the result into fibration basis in the last line. 
Higher-order terms in the $\eps$-expansion can be obtained without any problems. As an illustration, we have computed the expressions for $\mathscr{J}_{3,1}(\cQ)$ and $\mathscr{J}_{3,2}(\cQ)$ as well. The result is provided as ancillary material. This is only one of the six orthoschemes that contribute to the triangle integral in eq.~\eqref{eq:master}. The remaining five contributions are easy to obtain, because they can be obtained by a permutation of the external scales. As a result, we have obtained an algorithm to compute the one-loop triangle integral depending on six different scales to any desired order in $\eps$. 
As ancillary material, we provide a Mathematica file (\texttt{results.wl}) containing the explicit expressions for the functions \(\mathscr{J}_{3,i}\), \(J(t)\) and \(J(1-t)\).
%, where \texttt{sJ[3,i]} corresponds to \(\mathscr{J}_{3,i}\), \texttt{JFib[3,0]} to \(J(t)\) and \texttt{JFibInfty[3,0]} to \(J(1-t)\).\paulcomment{depending on how we defined J maybe a factor i/8 difference from results file to latex}\claudecomment{The Mathematica variables should be explained in the notebook, not here.}
In order to check our analytic result, we have evaluated it at several kinematic points in the Euclidean region using \texttt{GiNaC}~\cite{Bauer:2000cp,Vollinga:2004sn} and compared it to direct numerical integration of the Feynman-parameter representation using \texttt{NIntegrate} in Mathematica and to numerical evaluation using \texttt{pySecDec} \cite{Borowka:2017idc,Borowka:2018goh,Heinrich:2021dbf,Heinrich:2023til}, and we found very good agreement.
%\claudecomment{Comment on the checks you have made!}\paulcomment{Numerically checked against Feynman parameters using NIntegrate in Mathematica and against pySecDec (Euclidean); other regimes checked against pySecDec} 
We note that the expression here is valid in the Euclidean region, where the ordering is given by eq.~\eqref{eq:Euclidean_ordering}. We will discuss the analytic continuation to other kinematic regions in section~\ref{sec:analytic_continuation}.

%%%%%%%%%%%%%%%%%%%%%%%%%%%%%%%%%%%%%
%%%%%%%%%%%%%%%%%%%%%%%%%%%%%%%%%%%%%

\subsection{Results for box integrals}
\label{sec:boxes}

In this subsection we show how to obtain analytic results for the higher orders in $\eps$ for one-loop box integrals. We focus on integrals that are convergent for $\eps=0$. These are the most complicated integrals, because infrared divergences arise from internal and external masses being zero. For divergent box integrals closed expressions in terms of hypergeometric functions are often known (cf., e.g., refs.~\cite{Gehrmann:1999as,Brandhuber:2007yx}). We focus on the most general one-loop box integral depending on ten non-zero scales: the four propagator masses $m_i^2$ and six Mandelstam invariants $(l_i-l_j)^2$. 
The general strategy is very similar to the one outlined in section~\ref{sec:dimreg} and used for the computation of the one-loop triangle integral in the previous subsection. We will therefore be brief, and only highlight the main features that are specific to the case of the box integral. Complete results are provided in the ancillary file \texttt{results.wl}.

Our starting point is the projective dissection of the integral from section~\ref{sec:auxiliary-mass}. 
%The box integrals on the right-hand side of eq.~\eqref{eq:master} are then further decomposed into orthoschemes using the algorithms described in section~\ref{sec:hyperbolic}. 
We can then express the box integral as a sum of $4!=24$ volumes of orthoschemes. Each orthoscheme is defined by its Gram matrix of the form
\beq
\cQ = \begin{pmatrix}
\cQ_{1,1} & \cQ_{2,2} & \cQ_{3,3} & \cQ_{4,4} \\
\cQ_{2,2} & \cQ_{2,2} & \cQ_{3,3} & \cQ_{4,4} \\
\cQ_{3,3} & \cQ_{3,3} & \cQ_{3,3} & \cQ_{4,4} \\
\cQ_{4,4} & \cQ_{4,4} & \cQ_{4,4} & \cQ_{4,4} 
\end{pmatrix}\,.
\eeq
In the following we only discuss the contribution of a single orthoscheme with the choice \(\mu = \cQ_{4,4}\). All other contributions can be obtained in the same way. More concretely, we will be concerned with the computation of the integral
\beq\label{eq:curly_J_4}
\mathscr{J}_4(\cQ; \cQ_{4,4}) = \int_0^\infty\frac{\rd\eta}{\sqrt{1+\eta}}\,\eta^{-1-\eps}\,J_{4,0}^4(\cQ_{\eta})\,.
\eeq
For concreteness, we work in the Euclidean region where the entries of the Gram matrix are ordered as in eq.~\eqref{eq:Euclidean_ordering}. The analytic expression in terms of alternating MPLs for the function in the integrand can be obtained from ref.~\cite{Ren:2023tuj}:
\beq\label{eq:J4_box}
    J_{4,0}^4 (\cQ_\eta) = J_{4,0}^{4,(1)} (\cQ_\eta) + J_{4,0}^{4,(2)} (\cQ_\eta) %+ J_{4,0}^{4,(3)} (\cQ_\eta)
\eeq
with
\beq\bsp
    J_{4,0}^{4,(1)} (\cQ_\eta) &= 4 \operatorname{per}_\mathbb{R} \left[ \operatorname{ALi}_{1,1} \left( \frac{1+a \eta}{x_1^2}  ,y_1^2 \frac{1+\eta}{1+a\eta} \right) \right]\,, %\label{eq:box_2sqrt}
    \\
    J_{4,0}^{4,(2)} (\cQ_\eta) &= -  4\operatorname{per}_\mathbb{R} \left[ \operatorname{ALi}_{1,1} \left( \frac{1+\eta}{x_2^2}, y_2^2\right) \right]+ 4 \operatorname{per}_\mathbb{R} \left[\operatorname{ALi}_{1,1} \left( \frac{1+\eta}{x_3^2}, y_3^2\right) \right]\,,\label{eq:box_1sqrt}\
\esp\eeq
%s\claudecomment{I do not see why we need 3 terms. We only discuss $J_{4,0}^{4,(1)}$ separately. Would it not make more sense to combine $J_{4,0}^{4,(2)}$ and $J_{4,0}^{4,(3)}$?}\paulcomment{Yes we can do that. Would only add a sum sign in \eqref{eq:J4,0(2,3)}}
where we defined the ratio \(a = \tfrac{\cQ_{4,4}}{\cQ_{2,2}}>1\), and the inequality follows from the ordering in eq.~\eqref{eq:Euclidean_ordering} in the Euclidean region, and the arguments are
\begin{align}
\nonumber    x_1 &= \sqrt{\frac{\cQ_{2,2}-\cQ_{1,1}}{\cQ_{2,2}}}\,, && y_1 = \iu \sqrt{\frac{(\cQ_{3,3}-\cQ_{2,2}) \cQ_{4,4}}{\cQ_{2,2}(\cQ_{4,4}-\cQ_{3,3})}}\,,\\
    x_2 &= \sqrt{\frac{\cQ_{4,4}-\cQ_{1,1}}{\cQ_{4,4}}}\,, && y_2 = \iu \sqrt{\frac{(\cQ_{4,4} - \cQ_{1,1}) (\cQ_{3,3}-\cQ_{2,2})}{(\cQ_{2,2}-\cQ_{1,1})(\cQ_{4,4}-\cQ_{3,3})}}\,,\\
\nonumber    x_3 &= \sqrt{\frac{\cQ_{4,4}-\cQ_{3,3}}{\cQ_{4,4}}}\,, && y_3 = \iu \sqrt{\frac{\cQ_{3,3}-\cQ_{2,2}}{\cQ_{2,2}-\cQ_{1,1}}}\,.
\end{align}
%\paulcomment{\begin{equation}
%\begin{split}
%      J_{4,0}^4 (\cQ_\eta) = \frac{1}{2} \operatorname{per}_\mathbb{R} &\bigg[ \operatorname{ALi}_{1,1} \left( \frac{\cQ_{2,2}+\eta}{\cQ_{2,2}-\cQ_{1,1}},-\frac{(\cQ_{3,3}-\cQ_{2,2}) (\cQ_{4,4}+\eta)}{(\cQ_{2,2}+\eta)(\cQ_{4,4}-\cQ_{3,3})}\right) \\
%      &\quad -\operatorname{ALi}_{1,1} \left( \frac{\cQ_{4,4}+\eta}{\cQ_{4,4}-\cQ_{1,1}},-\frac{(\cQ_{4,4} - \cQ_{1,1}) (\cQ_{3,3}-\cQ_{2,2})}{(\cQ_{2,2}-\cQ_{1,1})(\cQ_{4,4}-\cQ_{3,3})}\right) \\
%      &\quad+\operatorname{ALi}_{1,1} \left( \frac{\cQ_{4,4}+\eta}{\cQ_{4,4}-\cQ_{3,3}},-\frac{\cQ_{3,3}-\cQ_{2,2}}{\cQ_{2,2}-\cQ_{1,1}}\right) \bigg].
%\end{split}
%\end{equation}
%Or do you want the full expressions with \(\perR\) expanded out?}
%From the discussion in section~\ref{sec:basic_simplex}, we know that this expression involves two distinct square roots of a linear polynomial in $\eta$. 
%We observe that the term \(J_{4,0}^{4,(1)} (\cQ_\eta) \) is the most complicated because it involves two distinct square roots while \(J_{4,0}^{4,(2)} (\cQ_\eta) \) and \(J_{4,0}^{4,(3)} (\cQ_\eta) \) only contain one. We will thus treat these two cases separately.
We introduce the \(\mathscr{J}_{4}^{(i)}(\cQ)\) as the contribution of \(J_{4,0}^{4,(i)} (\cQ_\eta)\) in eq.~\eqref{eq:J4_box} to the integral in eq.~\eqref{eq:curly_J_4}, i.e., we define
\beq
\mathscr{J}_4^{(i)}(\cQ; \cQ_{4,4}) = \int_0^\infty\frac{\rd\eta}{\sqrt{1+\eta}}\,\eta^{-1-\eps}\,J_{4,0}^{4,(i)}(\cQ_{\eta})\,.
\eeq
We have explicitly performed all integrations in terms of MPLs up to $\mathcal{O}(\eps^2)$. The final expressions are not particularly illuminating, and can be found in the ancillary files. We compared the analytic expression to a direct numerical integration of the corresponding Feynman-parameter integral, and we found good agreement.

We conclude this subsection by discussing some of the features that went into the evaluation of the integral.
The main technical difficulty in performing the integrations is the rationalisation of the square roots and the subtraction of the singularities at $\eta=0$. 
Note that not all the terms in eq.~\eqref{eq:J4_box} depend on both square roots. In fact, only \(J_{4,0}^{4,(1)} (\cQ_\eta) \) depends on both square-roots, while \(J_{4,0}^{4,(2)} (\cQ_\eta) \) 
%and \(J_{4,0}^{4,(3)} (\cQ_\eta) \) 
only depends on one of the two. In the remainder of this section, we discuss these two cases in turn.

\paragraph{Terms depending on both square roots.} 
Only \(J_{4,0}^{4,(1)} (\cQ_\eta) \) depends on both square roots simultaneously.
%\begin{equation}\label{eq:box_2sqrt}
%   X_{4,2}(\eta) := \operatorname{ALi}_{1,1}\!\! \left( \tfrac{\cQ_{22}+\eta}{\cQ_{22}-\cQ_{11}},\tfrac{(\cQ_{22}-\cQ_{33}) (\cQ_{44}+\eta)}{(\cQ_{22}+\eta)(\cQ_{44}-\cQ_{33})}\right)\,.
%\end{equation}
The square roots in question are \(\sqrt{1+\eta}\) and \(\sqrt{1+a \eta}\) (note that \(\sqrt{1+\eta}\) also appears in the integrand). 
%After factoring out \(\cQ_{22}\) and \(\cQ_{44}\) and rescaling $\eta\to\cQ_{44}\eta$, we are left with the square roots \(\sqrt{1+\eta}\) and \(\sqrt{1+a \eta}\) where \(a= \tfrac{\cQ_{44}}{\cQ_{22}}>1\), and the inequality follows from the ordering in eq.~\eqref{eq:Euclidean_ordering} in the Euclidean region. 
%
We rationalise these square roots in two steps. First, we perform the change of variables $\eta =  t^2-1$, which rationalises one of them,
\begin{equation}
    \begin{split}
        \sqrt{1+\eta} & = t\,,\\
        \sqrt{1+a \eta} &= \sqrt{1-a + a t^2 } = \sqrt{a-1} \ \sqrt{\left(\tfrac{t}{A}\right)\!\!\!\!\!\!\!\!\!\!\!\!{\phantom{\left(\tfrac{t}{A}\right)}}^2-1}\,,
    \end{split}
\end{equation}
where we defined \(A=\sqrt{\frac{a-1}{a}}<1\). In a second step we rationalise the remaining square root through the change of variables $t=A\,\frac{1+\tau^2}{1-\tau^2}$, such that 
\beq
\sqrt{\left(\tfrac{t}{A}\right)\!\!\!\!\!\!\!\!\!\!\!\!{\phantom{\left(\tfrac{t}{A}\right)}}^2-1} = \frac{2\,\tau}{1-\tau^2}\,.
\eeq
The final integration range for $\tau$ is $[r,1]$, with $r=\sqrt{\tfrac{1-A}{1+A}}<1$. For practical purposes it turns out to be convenient to perform an affine shift $\tau=(1-r)v+r$ that maps the final integration region to the segment $[0,1]$.
%
%We define the kinematic variables
%\begin{equation}
%    x = \sqrt{\frac{\cQ_{22}-\cQ_{11}}{\cQ_{22}}}, \qquad y = \sqrt{\frac{(\cQ_{33}-\cQ_{22}) \cQ_{44}}{\cQ_{22}(\cQ_{44}-\cQ_{33})}}
%\end{equation}
%\claudecomment{Following unchanged. Again, I don't understand how you can rescale kinematic variables...}
In total we get the following transformations,
\begin{equation}\bsp
    \sqrt{1+a \eta} &= \frac{\frac{(1-r) v }{r}+1}{(1-v ) \left(\frac{(1-r) v }{r+1}+1\right)}\,,\\[1.25ex]
    \sqrt{\frac{1+\eta}{1+a\, \eta}} &= \frac{\left(1+\frac{(1-r) v }{r-\iu}\right) \left(1+\frac{(1-r) v }{r+\iu}\right)}{\frac{(1-r) v }{r}+1}\,,
\esp\end{equation}
which we can insert into \(J_{4,0}^{4,(1)} (\cQ_\eta) \). After passing to a fibration basis, we get: %\claudecomment{The brackets and the $\tfrac{i}{8}$ factors look weird...}
%\claudecomment{What is $J(v)$? It is not defined!}\paulcomment{Replace \(J(v) = \perR X_{4,2} (\eta (v))\) need to fix the constants}
\begin{align}
\nonumber    \hat{J}_{4,0}^{4,(1)} (\cQ_v)
    &= \hat{J}_{4,0}^{4,(1)} (\cQ) + \sum_{s_1,s_2\in \{-1,1\}} \bigg\{ G(c_{s_1},b_{s_2}; v ) - G(c_{s_1},b'_{s_2}; v) - G(b'_{s_1},b_{s_2};v) \bigg\} \\
 \label{eq:Box_v-fibre}   &\quad+\sum_{s_1\in \{-1,1\}} \bigg\{ G(1;x_1) \left[ G(c_{s_1}; v) - G(b'_{s_1}; v) \right]  - G(1;y_1) G(c_{s_1};v) \bigg\} \,,
%\nonumber    \hat{J}_{4,0}^{4,(1)} (\cQ_v)
%    &= \sum_{s_1,s_2\in \{-1,1\}} \bigg\{ G (a_{s_1},b'_{s_2};v)+ G(c_{s_1},b_{s_2}; v ) - G(c_{s_1},b'_{s_2}; v) - G(b'_{s_1},b_{s_2};v) \bigg\} \\
% \label{eq:Box_v-fibre}   &\quad+\sum_{s_1\in \{-1,1\}} \bigg\{ G(1;x) \left[ G(c_{s_1}; v) - G(b'_{s_1}; v) \right] + G\! \left(b'_{s_1},-\tfrac{r}{1-r};v \right )\\
% \nonumber   &\quad  + G(1;y) \left[ G( a_{s_1} ;v )  - G(c_{s_1};v)\right] - G\!\left( a_{s_1} , -\tfrac{r}{1-r} ; v \right) \bigg\} + \hat{J}_{4,0}^{4,(1)} (\cQ)\,,
\end{align}
%Sum[(G[1, x] - G[1, y]) G[c[s1], tau] - 
%   G[1, x] G[bp[s1], tau], {s1, {-1, 1}}] + 
% Sum[G[c[s1], b[s2], tau] - G[c[s1], bp[s2], tau] - 
%   G[bp[s1], b[s2], tau], {s1, {-1, 1}}, {s2, {-1, 1}}]
where we defined
\begin{equation}\bsp
    %a_\pm &= \frac{\pm \iu - r}{1-r}\,,\\
    b_\pm &= -\frac{1-r^2+ 2r^2 x_1 \pm \sqrt{(1-r^2)^2+r^2 (2x_1)^2}}{2 (1-r) rx_1}\,,\\
    b'_\pm &= \frac{1+r^2 - 2r^2 y_1 \pm \sqrt{(1+r^2)^2-r^2 (2y_1)^2}}{2 (1-r) ry_1}\,,\\
    c_\pm &= -\frac{r^3 (x_1-y_1) +r(x_1+y_1) \pm \sqrt{(1+r^2)^2 x_1^2-(1-r^2)^2 y_1^2}}{(1-r) \left( (1+r^2) x_1 + (1-r^2)y_1 \right)}\,,
\esp\end{equation}
and \( \hat{J}_{4,0}^{4,(1)} (\cQ) = [G(1; x_1) - G(1; y_1)] G(y_1; x_1) - G(1, y_1; x_1)\) in analogy to eq.~\eqref{eq:tri_eps=0}.
%\((\rt_+ \to \rt_-)\) means %an analogous thing to \((\sqrt{\phantom{.}} \to -\sqrt{\phantom{.}})\)
%to sum over all possible roots, e.g., \claudecomment{This is not good, it leads to confusion and misinterpretation.}
%\begin{equation}
%    \begin{split}
%    &\qquad G ( \rt_+ (\iu), \rt_+(y);v ) + (\rt_+ \to \rt_-) \\
%    &= G ( \rt_+ (\iu), \rt_+(y);v ) +G ( \rt_+ (\iu), \rt_-(y);v )+G ( \rt_- (\iu), \rt_+(y);v ) \\
%    &\qquad +G ( \rt_- (\iu), \rt_-(y);v ) \,    .  
%    \end{split}
%\end{equation}

So far we have discussed how to rationalise all square roots that appear in the quantity \(J_{4,0}^{4,(1)} (\cQ_v)\), and we have shown that after rationalision, we can pass to a fibration basis. The last step is to subtract the singularities at $\eta=0$, so that we can expand in $\eps$ under the integral sign.
After substituting all the changes of variables and rearranging terms, we find
\begin{equation}
    \begin{split}
        \mathscr{J}_{4}^{(1)}(\cQ) &= \int_0^1 \Df v\ \left( \frac{1}{v} + \frac{1}{v+\frac{2r}{1-r}} -\frac{1}{v+\frac{1+r^2}{r(1-r)}}-\frac{1}{v-1-\frac{1}{r}} \right) F(v)^{-\varepsilon} J_{4,0}^{4,(1)} (\cQ_v) \\ 
        &= -\frac{1}{\varepsilon} F_0^{-\varepsilon} J_{4,0}^{4,(1)} (\cQ) + \int_0^1 \frac{\Df v}{v}\ \left[ F(v)^{-\varepsilon} J_{4,0}^{4,(1)} (\cQ_v) - (v\, F_0)^{-\varepsilon} J_{4,0}^{4,(1)} (\cQ) \right]\\
        &\qquad + \int_0^1 \Df v\ \left( \frac{1}{v+\frac{2r}{1-r}} -\frac{1}{v+\frac{1+r^2}{r(1-r)}}-\frac{1}{v-1-\frac{1}{r}} \right) F(v)^{-\varepsilon} J_{4,0}^{4,(1)} (\cQ_v)\,,
    \end{split}
\end{equation}
%\claudecomment{What are these quantities? Neither $\mathscr{J}^{(1)}_\Box(\varepsilon)$ not $J(v)$ was defined! We should show everything in relation to $X_{4,2}$!}\paulcomment{Would be \(\mathscr{J}_{4,0} (\cQ)\) but only including the summand from \(X_{4,2}\) since the other 2 summands are treated separately}
where we defined the quantities
\beq\bsp
F(v) &\,=  \frac{4 r^2}{(1+r^2)^2}\ \frac{v \left(\frac{2 r}{1-r}+ v\right) \left( \frac{1+r}{r}-v \right)\left( \frac{1+r^2}{r(1-r)}+v \right)}{(1-v)^2 \left( \frac{1+r}{1-r}+v \right)^2}\,,\\
F_0 &\,= \lim_{v\to 0}\frac{F(v)}{v} = \frac{8 r}{1+r +r^2+ r^3}\,.
\esp\eeq
All the integrals are absolutely convergent, and we may expand in $\eps$ under the integral sign. The resulting integral can easily be performed via the definition of MPLs in eq.~\eqref{eq:MPL_def}.

\paragraph{Terms depending on one square-root.} %\claudecomment{I just copied this over. This needs to be rewritten, it is too intransparent! You have 3 different definitions of $x$ and $y$ at this point!!}
Let us now consider the contribution from \(J_{4,0}^{4,(2)} (\cQ_\eta)\).
%and \(J_{4,0}^{4,(3)} (\cQ_\eta)\).
%\begin{gather}
%    -\operatorname{ALi}_{1,1} \left( \frac{\cQ_{4,4}+\eta}{\cQ_{4,4}-\cQ_{1,1}},-\frac{(\cQ_{4,4} - \cQ_{1,1}) (\cQ_{3,3}-\cQ_{2,2})}{(\cQ_{2,2}-\cQ_{1,1})(\cQ_{4,4}-\cQ_{3,3})}\right)\label{eq:Box_one_sqrt_2} \\
%     \operatorname{ALi}_{1,1} \left( \frac{\cQ_{4,4}+\eta}{\cQ_{4,4}-\cQ_{3,3}},-\frac{\cQ_{3,3}-\cQ_{2,2}}{\cQ_{2,2}-\cQ_{1,1}}\right) \label{eq:Box_one_sqrt_3} 
%\end{gather}
%The general strategy for both expressions is similar, because they only involve the square root \(\sqrt{1+\eta}\). 
The only square root present is \(\sqrt{1+\eta}\). Hence, the transformation we need is the same as in the case for the triangle integral in eq.~\eqref{eq:Tri_t-trafo}, i.e., we have
\begin{equation}
    \sqrt{1+\eta} = \frac{1}{1-t}\,.
\end{equation}
%where \(x\) and \(y\) are
%\begin{equation}
%    x=\sqrt{\frac{\cQ_{4,4}-\cQ_{1,1}}{\cQ_{4,4}}}, \qquad y= \iu\ \sqrt{\frac{(\cQ_{4,4} - \cQ_{1,1}) (\cQ_{3,3}-\cQ_{2,2})}{(\cQ_{2,2}-\cQ_{1,1})(\cQ_{4,4}-\cQ_{3,3})}}
%\end{equation}
%for \eqref{eq:Box_one_sqrt_2} or
%\begin{equation}
%    x=\sqrt{\frac{\cQ_{4,4}-\cQ_{3,3}}{\cQ_{4,4}}}, \qquad y= \iu\ \sqrt{\frac{\cQ_{3,3}-\cQ_{2,2}}{\cQ_{2,2}-\cQ_{1,1}}}.
%\end{equation}
%for \eqref{eq:Box_one_sqrt_3}.
%where \((x,y) = \begin{pmatrix} \frac{\Q_{4,4}-\Q_{1,1}}{\Q_{4,4}} & -\frac{(\cQ_{4,4} - \cQ_{1,1}) (\cQ_{3,3}-\cQ_{2,2})}{(\cQ_{2,2}-\cQ_{1,1})(\cQ_{4,4}-\cQ_{3,3})} \end{pmatrix}\)
%or \((x,y) =  \begin{pmatrix}\frac{\cQ_{4,4}+\Q_{3,3}}{\cQ_{4,4}} & -\frac{\cQ_{3,3}-\cQ_{2,2}}{\cQ_{2,2}-\cQ_{1,1}} \end{pmatrix}\) respectively.
In these coordinates we obtain the following expression in a fibration basis:
\begin{align}
 \nonumber   \hat{J}_{4,0}^{4,(2)} (\cQ_t) &=\sum_{i\in\{2,3\}} (-1)^{i+1} \bigg\{ G \left( 1- \frac{y_i}{x_i}, - \frac{1-x_i}{x_i};t \right)  + \left[ G(1;x_i)- G(1;y_i) \right] G \left( 1- \frac{y_i}{x_i};t\right) \bigg\}\\
 \label{eq:J4,0(2,3)}   &\quad+ \hat{J}_{4,0}^{4,(2)} (\cQ) \,,
\end{align}
where \(\hat{J}_{4,0}^{4,(2)} (\cQ) = \sum_{i\in\{2,3\}} \left( (-1)^{i+1} [G(1; x_i) - G(1; y_i)] G(y_i; x_i) - G(1, y_i; x_i) \right)\) is the constant part at \(t=0\).

In contrast to the triangle integral, the integration kernel is different due to the contribution from the Jacobian
\begin{equation}
    \frac{\Df{\eta}}{\eta\ \sqrt{1+\eta}} =\Df t\ \left( \frac{1}{t} + \frac{1}{2-t} \right)\,.
\end{equation}
%Notice that there is no longer a (spurious) divergences at \(t \to 1\) (which corresponds to \(\eta \to \infty\)). 
%
%From all these expressions it is now an easy task to obtain the result. 
If we let
\begin{equation}
    h(t) = \frac{t(2-t)}{(1-t)^2}\,,
\end{equation}
then we obtain the following integral:
\beq\bsp
    \mathscr{J}_{4}^{(2)}(\cQ) &= \int_0^1 \Df t\ \left( \frac{1}{t} + \frac{1}{2+t} \right) h(t)^{-\varepsilon} J^{4,(2)}_{4,0} (\cQ_t) \\ 
    &= -\frac{1}{\varepsilon} 2^{-\varepsilon} J^{4,(2)}_{4,0} (\cQ) + \int_0^1 \frac{\Df t}{t}\ \left( h(t)^{-\varepsilon} J^{4,(2)}_{4,0} (\cQ_t) - (2 t)^{-\varepsilon} J^{4,(2)}_{4,0} (\cQ) \right) \\
    &\qquad+ \int_0^1 \frac{\Df t}{2-t}\ h(t)^{-\varepsilon} J^{4,(2)}_{4,0} (\cQ_t)\,.
\esp\eeq
The remaining integrals are finite as $\eps\to0$ and can be expanded under the integration sign.
%Here \(\mathscr{J}^{(2)}\) corresponds to \eqref{eq:Box_one_sqrt_2} while \(\mathscr{J}^{(3)}\) corresponds to \eqref{eq:Box_one_sqrt_3}.
%
%The final result for the box can then be written in similar form as \eqref{eq:TriRes}. The only difference is that we have more summands:
%\begin{equation}\label{eq:BoxRes}
%\begin{gathered}
%    \sqrt{\det{Q}}\ I_4^{4-2\varepsilon} = \frac{1}{2} \Omega_{-2 \varepsilon} \pi^{\varepsilon} Q_{4,4}^{-\varepsilon} \mathscr{J}(\varepsilon),\\
%    \textnormal{where } \mathscr{J}(\varepsilon) = \mathscr{J}^{(1)}(\varepsilon)-\mathscr{J}^{(2)}(\varepsilon)+\mathscr{J}^{(3)}(\varepsilon).
    %&= J(0) - \left( \mathscr{J}_\textnormal{red} (0) + \log Q_{4,4} \  J(0) \right) \varepsilon \\
    %&\qquad - \left( \mathscr{J}^\prime_\textnormal{red} (0) + \log Q_{4,4} \mathscr{J}_\textnormal{red} (0) + \frac{\log^2 Q_{4,4}}{2} J(0) \right) \varepsilon^2 + \mathcal{O}(\varepsilon^3)
%\end{gathered}
%\end{equation}

%\paragraph{Conclusion.} 
%\claudecomment{Comment on the checks you have made!}

%%%%%%%%%%%%%%%%%%%%%%%%%%%%%%%%%%%%%%
%%%%%%%%%%%%%%%%%%%%%%%%%%%%%%%%%%%%%%

\subsection{Results for pentagon integrals}
\label{sec:pentagon}
%\paulcomment{@paul check prefactors}
%
The evaluation of the pentagon integral at higher orders in \(\varepsilon\) is similar to the triangle and box integrals. This is due to the fact that the integration kernel is similar to the triangle and there are at most two distinct square-roots as for the box  integral. The main difference is that the pentagon integral at \(\varepsilon=0\) is a function of weight three, which makes the computation lengthier.
%--already in Euclidean kinematics we have to differentiate three cases because three arguments can form non-trivial combinations. 
We will therefore be brief, and only highlight the main steps. The results up to order $\mathcal{O}(\eps^1)$  can be found in the ancillary file \texttt{results.wl}.
%\claudecomment{Need ot provide filename!}

Let us begin by recalling  the volume of the hyperbolic simplex attached to the pentagon integral. We can obtain this volume via the same limiting procedure we used for the triangle result at \(\varepsilon=0\). Using \(\mu = \cQ_{4,4}\) throughout this section, we have:
\beq\label{eq:J6_Penta}
    J_{5,0}^6 (\cQ_\eta) = -J_{5,0}^{6,(1)} (\cQ_\eta) + J_{5,0}^{6,(2)} (\cQ_\eta) + J_{5,0}^{6,(3)} (\cQ_\eta)\,,
\eeq
%\beq\bsp\label{eq:J6_Penta}
%    J_{5,0}^6 (\cQ_\eta) &= -J_{5,0}^{6,(1)} (\cQ_\eta) + J_{5,0}^{6,(2)} (\cQ_\eta) - J_{5,0}^{6,(3)} (\cQ_\eta) + J_{5,0}^{6,(4)} (\cQ_\eta) - J_{5,0}^{6,(5)} (\cQ_\eta) \\
%    &\quad + J_{5,0}^{6,(6)} (\cQ_\eta) - J_{5,0}^{6,(7)} (\cQ_\eta) + J_{5,0}^{6,(8)} (\cQ_\eta) - J_{5,0}^{6,(9)} (\cQ_\eta)\,,
%\esp\eeq
with 
\begin{align}\label{eq:penta_1sqrt2}
 \nonumber   J_{5,0}^{6,(1)} (\cQ_\eta) &= 4 \operatorname{per}_\mathbb{R} \left[ \operatorname{ALi}_{1,1,1} \left( \frac{1+a \eta}{x_1^2}  ,y_1^2 \frac{1+\eta}{1+a\eta}, \frac{ z_1^2 }{1+\eta}  \right) \right]\,,\\
  \nonumber  J_{5,0}^{6,(2)} (\cQ_\eta) &= 4 \sum_{i \in \{2,3\}} (-1)^{i} \operatorname{per}_\mathbb{R} \left[ \operatorname{ALi}_{1,1,1} \left( \frac{1+\eta}{x_i^2}, \frac{y_i^2}{1+\eta}, z_i^2\right) \right] \\
  &\quad + 4 \sum_{j \in \{ 4,5 \}} (-1)^j \operatorname{per}_\mathbb{R} \left[\operatorname{ALi}_{1,1,1} \left( \frac{1+\eta}{x_j^2}, y_j^2, \frac{z_j^2}{1+\eta}\right) \right] \\
\nonumber  &\quad + 4 \sum_{w \in \{8,9\}} (-1)^w \operatorname{per}_\mathbb{R} \left[\operatorname{ALi}_{1,2} \left( \frac{1+\eta}{x_w^2}, {x'}_w^2\right) \right] \,,\\
 \nonumber    J_{5,0}^{6,(3)} (\cQ_\eta) &= 4 \sum_{k \in \{6,7\}} (-1)^k \operatorname{per}_\mathbb{R} \left[\operatorname{ALi}_{1,1,1} \left( \frac{1+a\eta}{x_k^2}, \frac{y_k^2}{1+a\eta},z_k^2\right) \right]\,,
\end{align}
%\claudecomment{Same comment as before: Would it not make more sense to only define the sum of $ J_{5,0}^{6,(k)}$ with $k\ge 2$?}\paulcomment{We can sum 2,3,4,5,8,9. The contributions of 6,7 require one more change or variables and thus an additional prefactor \(a^\eps\)}
%\begin{align}
% \nonumber   J_{5,0}^{6,(1)} (\cQ_\eta) &= 4 \operatorname{per}_\mathbb{R} \left[ \operatorname{ALi}_{1,1,1} \left( \frac{1+a \eta}{x_1^2}  ,y_1^2 \frac{1+\eta}{1+a\eta}, \frac{ z_1^2 }{1+\eta}  \right) \right]\,,\\
%  \nonumber  J_{5,0}^{6,(i)} (\cQ_\eta) &= 4 \operatorname{per}_\mathbb{R} \left[ \operatorname{ALi}_{1,1,1} \left( \frac{1+\eta}{x_i^2}, \frac{y_i^2}{1+\eta}, z_i^2\right) \right]\,, && i \in \{2,3\}\,,\\
%    J_{5,0}^{6,(j)} (\cQ_\eta) &= 4 \operatorname{per}_\mathbb{R} \left[\operatorname{ALi}_{1,1,1} \left( \frac{1+\eta}{x_j^2}, y_j^2, \frac{z_j^2}{1+\eta}\right) \right]\,, && j \in \{4,5\}\label{eq:penta_1sqrt2}\,,\\
% \nonumber    J_{5,0}^{6,(k)} (\cQ_\eta) &= 4 \operatorname{per}_\mathbb{R} \left[\operatorname{ALi}_{1,1,1} \left( \frac{1+a\eta}{x_k^2}, \frac{y_k^2}{1+a\eta},z_k^2\right) \right]\,, && k \in \{6,7\}\,,\\
%  \nonumber   J_{5,0}^{6,(w)} (\cQ_\eta) &= 4 \operatorname{per}_\mathbb{R} \left[\operatorname{ALi}_{1,2} \left( \frac{1+\eta}{x_w^2}, {x'}_w^2\right) \right]\,, && w \in \{8,9\}\,,
%\end{align}
and arguments:
%\allowdisplaybreaks
\begin{align}
  \nonumber  x_1 &= \sqrt{\frac{\cQ_{2,2}-\cQ_{1,1}}{\cQ_{2,2}}}\,, && y_1 = \frac{y_6}{x_3}\,, && z_1 = \iu \sqrt{\frac{\cQ_{5,5}-\cQ_{4,4}}{\cQ_{4,4}}} \,,\\
\nonumber    x_2 &= \sqrt{\frac{\cQ_{4,4}-\cQ_{1,1}}{\cQ_{4,4}}}\,, && y_2 = z_1\,, && z_2 = \frac{y_6 x_2}{x_1 x_3}\,,\\
 \nonumber   x_3 &= \sqrt{\frac{\cQ_{4,4}-\cQ_{3,3}}{\cQ_{4,4}}}\,, && y_3 = z_1\,,&& z_3 = \frac{y_6}{x_1}\,,\\
\nonumber    x_4 &= x_2\,, && y_4 = z_2 \,,&& z_4 = z_1\,,\\
    x_5 &= x_3\,, && y_5 = z_3\,,&& z_5 = z_1\,,\\
\nonumber    x_6 &= x_1\,, && y_6 =  \iu \sqrt{\frac{\cQ_{3,3}-\cQ_{2,2}}{\cQ_{2,2}}}\,,&& z_6 = \frac{z_1}{x_3}\,,\\
\nonumber    x_7 &= x_1\,, && y_7 =  \iu \sqrt{\frac{\cQ_{5,5}-\cQ_{2,2}}{\cQ_{2,2}}}\,,&& z_7 = \frac{y_6 z_1}{y_7 x_3}\,,\\
\nonumber    x_8 &= x_2\,, && x'_8 = \frac{z_1 y_6 x_2}{x_1 x_3} \,,\\
\nonumber    x_9 &= x_3\,, && x'_9 = \frac{z_1 y_6}{x_1} \,.
\end{align}
Notice that the variables \(x_1,x_2,x_3,y_6,y_7,z_1\) are not independent. We have the relations
\beq\bsp
y_6 &\,= \sqrt{\tfrac{ (1-x_3^2) x_1^2 -x_2^2+x_3^2 }{1-x_2^2}}\,,\\
y_7 &\,= \sqrt{\tfrac{ (1 - z_1^2) x_1^2 -x_2^2 + z_1^2 }{1-x_2^2}}\,.
\esp\eeq
However, we find it convenient to keep these variables independent throughout the computation. 

%\claudecomment{I do not understand what you saying here...}
%\paulcomment{Since \(J_{5,0}^{6,(3)} (\cQ_\eta)\) involves \(\sqrt{1+a\eta}\) and not \(\sqrt{1+\eta}\) so we need one more change of variable in the integrand of \(\int_0^\infty \mathrm{d}\eta \, \eta^{-1-\eps} J_{5,0}^{6,(3)} (\cQ_\eta) = a^\eps \int_0^
%\infty\mathrm{d}\eta \, \eta^{-1-\eps} J_{5,0}^{6,(3)} (\cQ_{\tfrac{\eta}{a}})  \). \(a^\eps J_{5,0}^{6,(3)} (\cQ_{\tfrac{\eta}{a}})\) now only involves \(\sqrt{1+\eta}\) and we can treat \(J_{5,0}^{6,(3)} (\cQ_\eta)\) and \(a^\eps J_{5,0}^{6,(3)} (\cQ_{\tfrac{\eta}{a}})\) on the same footing; hence, we might combine them into \({J'}_{5,0}^{6,(2)} (\cQ_\eta)\)}\paulcomment{better?}
The function \(J_{5,0}^{6,(3)} (\cQ_\eta)\) contains a square root of the form \(\sqrt{1+a\eta}\). This is slightly different from the square root appearing in \(J_{5,0}^{6,(2)} (\cQ_\eta)\), which depends on \(\sqrt{1+\eta}\). To place both terms on the same footing and to be able to combine them, we perform a simple rescaling under the auxiliary mass integral:
\beq
    \int_0^\infty \mathrm{d} \eta \, \eta^{-1-\eps} J_{5,0}^{6,(3)} (\cQ_\eta) = a^\eps \int_0^\infty \mathrm{d}\eta\, \eta^{-1-\eps} J_{5,0}^{6,(3)} \big(\cQ_{\frac{\eta}{a}}\big)\,.
\eeq
After this rescaling, the integrand \(a^\eps J_{5,0}^{6,(3)} \big(\cQ_{\frac{\eta}{a}}\big)\) depends on \(\sqrt{1+\eta}\) instead of \(\sqrt{1+a\eta}\). Therefore, both terms \(J_{5,0}^{6,(2)} (\cQ_\eta)\) and \(a^\eps J_{5,0}^{6,(3)} \big(\cQ_{\frac{\eta}{a}}\big)\) now have an identical square root structure. This allows us to treat them on the same footing.

We now briefly highlight the main features of the computation of the contributions from the different terms in eq.~\eqref{eq:penta_1sqrt2}, in particular how we rationalise the square-roots and how we subtract the divergences at $\eta=0$.
We start by discussing the contribution from \(J_{5,0}^{6,(1)} (\cQ_\eta)\), which is the only contribution that depends on both square roots \(\sqrt{1+\eta}\) and \(\sqrt{1+a\eta}\). We already know from the discussion of the box integral in section~\ref{sec:boxes} how to rationalise these square roots. The only difference to the case of the box integral is the absence of \(\sqrt{1+\eta}\) in the denominator of the integrand, which only appears for integrals with $N$ even. We therefore get a slightly different integration kernel when changing coordinates from \(\eta\) to \(v\):
\beq
    \frac{\Df{\eta}}{\eta} =\Df v\ \left( \frac{1}{v} + \frac{2}{1-v}+ \frac{1}{v+\frac{2r}{1-r}} +\frac{1}{v+\frac{1+r^2}{r(1-r)}}+\frac{1}{v-1-\frac{1}{r}}  -\frac{2}{v+\frac{1+r}{1-r}}\right)\,.\label{eq:penta_v_ker}
\eeq
Similarly to before, we have
\begin{equation}
    \begin{split}
        \mathscr{J}_5^{(1)}(\cQ) &= \int_0^1 \Df v\ \left[\frac{1}{v} + \frac{2}{1-v} + R(v) \right] F(v)^{-\varepsilon} J_{5,0}^{6,(1)} (\cQ_v) \\ 
        &= -\frac{1}{\varepsilon} F_0^{-\varepsilon} J(0) + \int_0^1 \frac{\Df v}{v}\ \left[ F(v)^{-\varepsilon} J_{5,0}^{6,(1)} (\cQ_v) - (v\, F_0)^{-\varepsilon} J(0) \right]\\
        &\qquad + \int_0^1 \frac{2\ \Df v}{1-v}\ F(v)^{-\varepsilon} J_{5,0}^{6,(1)} (\cQ_v) + \int_0^1 \Df v\ R(v) F(v)^{-\varepsilon} J_{5,0}^{6,(1)} (\cQ_v)\,,
    \end{split}
\end{equation}
with
\begin{equation}
    R(v) \coloneqq \frac{1}{v+\frac{2r}{1-r}} +\frac{1}{v+\frac{1+r^2}{r(1-r)}}+\frac{1}{v-1-\frac{1}{r}}  -\frac{2}{v+\frac{1+r}{1-r}}\,.
\end{equation}
On the other hand, all the remaining terms are, by construction, functions only of the square root \(\sqrt{1+\eta}\).
%Namely, for \(i \in \{2,3,4,5,8,9\}\) only \(\sqrt{1+\eta}\) appears, while for \(i\in \{6,7\}\) only \(\sqrt{1+a \eta}\) appears. 
We can thus rationalise these square roots using the change of variables already used for the evaluation of the triangle integral in section~\ref{sec:triangles}. 

\section{Analytic continuation}\label{sec:analytic_continuation}

Our results for triangle, box and pentagon integrals from the previous section are valid in the Euclidean region, where all Mandelstam invariants are negative and the integrals are real. In applications it is typically required to analytically continue the results to a scattering region where some of the Mandelstam invariants are positive, and the integrals develop a non-trivial imaginary part. At the level of the analytic expressions, this corresponds for example to finding a representation in terms of MPLs that individually admit a convergent series representation as in eq.~\eqref{eq:Li_def}. In practise, this can be a complicated endeavour, because the MPLs may have a complicated algebraic dependence on the kinematic variables. In this section we discuss how we can use the representation of one-loop integrals as a combination of volumes of orthoschemes to perform the analytic continuation to other kinematic regions. While our procedure is general and can be applied to all integrals discussed in this paper, we illustrate it on the example of the one-loop triangle depending on six independent kinematic variables. We will, however, often keep the dependence on $N$ generic.

Our starting point is the representation of a one-loop integral obtained from the projective dissection in eq.~\eqref{eq:JdN0_proj_diss}. The main point is that the individual terms in eq.~\eqref{eq:JdN0_proj_diss} themselves admit an interpretation in terms of Feynman integrals, but in special kinematics corresponding to the Gram matrix $\cQ$ of an orthoscheme from the projective dissection (cf.~eq.~\eqref{eq:Gram_special}). The idea is then to separately continue the contribution from each orthoscheme in eq.~\eqref{eq:JdN0_proj_diss}. Note that each orthoscheme depends on fewer dimensionless variables (\(N-1\) in total) than the full Feynman integral in general kinematics. For example, the orthoscheme contributing to the triangle integral depend on only two dimensionless ratios, while the complete integral depends on five dimensionless ratios.  %This means we need to analytically continue a factorially-growing number of functions if we want to cover all kinematic regions. 

Let us now focus on an orthoscheme with Gram matrix $\cQ$, and let us discuss the general strategy for its analytic continuation. The alternating MPLs that appear in the contribution from this orthoscheme are only functions of the entries of $\cQ$, and we know that in the Euclidean region the entries of $\cQ$ are real and ordered according to eq.~\eqref{eq:Euclidean_ordering}. Different regions then correspond to kinematic configurations where this ordering is violated. The boundaries between the different kinematic regions are also easily described, and correspond to points where $\cQ_{i,i}=\cQ_{i+1,i+1}$, for some $i$. It is possible to work out the analytic continuation of the alternating MPLs as a function of the entries of $\cQ$. This, however, introduces ambiguities, because the sign of the terms proportional to $2\pi i$ are fixed by the $i0$ prescription of the Mandelstam invariants,
\beq
\log(-(l_i-l_j)^2-i0) = \log|(l_i-l_j)^2| - i\pi\,\theta(-(l_i-l_j)^2)\,,
\eeq
where $\theta$ denotes the Heaviside step function. Here our strategy is to fix this ambiguity by comparing numerical results in a  given region obtained from our analytic expressions to a direct numerical evaluation in \texttt{pySecDec}.\footnote{A more systematic approach to determine the analytic continuation of Feynman integrals using the coproduct of MPLs is laid out in ref.~\cite{Duhr:2015rjo}.} Note that this is possible because in the projective dissection, each term in the decomposition is itself a Feynman integral. We stress that one only needs a finite number of numerical evaluations to fix all the ambiguities, and after that the analytic expression in a given region is uniquely fixed.

%\subsection*{Strategy}
%The analytic continuation proceeds by considering all possible orderings of the Gram matrix entries, analogously to the Euclidean ordering in eq.~\eqref{eq:Euclidean_ordering}. Starting in the Euclidean region, we vary the entries of \(\cQ\) until two adjacent diagonal elements \(\cQ_{i,i}\) and \(\cQ_{i+1,i+1}\) cross, i.e.,:
%\beq
%    0 < \mathcal{Q}_{11} < \ldots < \mathcal{Q}_{i-1,i-1} < \mathcal{Q}_{i+1,i+1} < \mathcal{Q}_{i,i} < \mathcal{Q}_{i+2,i+2} < \ldots < \mathcal{Q}_{N,N}\,.
%\eeq
%Only at such crossings can non-trivial analytic structure, such as discontinuities or imaginary parts in \(I^D_N(\cQ)\), arise. This is because the crossings are precisely the points at which the arguments of the alternating polylogarithms may switch between different domains (e.g.\ \(<1\), \(>1\), or purely imaginary), thereby triggering changes in their branch structure.
%The general idea is to determine the continuation across a given crossing through enumerating all possible imaginary contributions coming from branch cuts along the path and then fixing the correct branches by comparison with numerical evaluation using \texttt{pySecDec}. Proceeding iteratively, we determine the analytic form in each region.

%\subsection*{Counting regions}

%\subsection*{Triangle case}

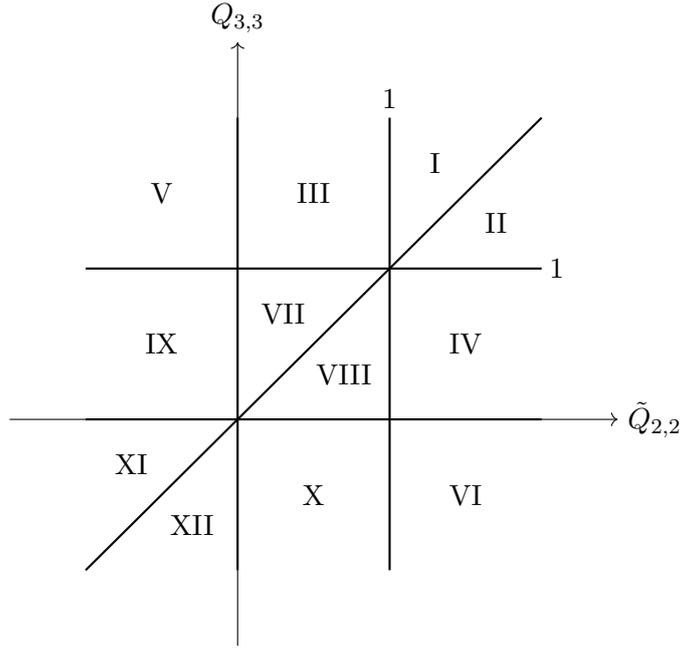
\begin{figure}[h!]
    \centering
    \begin{tikzpicture}[scale=2]
        
        % Draw axes
        \draw[->] (-1.5,0) -- (2.5,0) node[right] {$\tilde{Q}_{2,2}$};
        \draw[->] (0,-1.5) -- (0,2.5) node[above] {$\tilde{Q}_{3,3}$};
        
        % Draw crosses at (0,0) and (1,1)
        \draw[thick] (-1,0) -- (2,0); % Horizontal part of cross at (0,0)
        \draw[thick] (0,-1) -- (0,2); % Vertical part of cross at (0,0)
        \draw[thick] (-1,1) -- (2,1); % Horizontal part of cross at (1,1)
        \draw[thick] (1,-1) -- (1,2); % Vertical part of cross at (1,1)
        
        % Draw the diagonal line (f(x) = x)
        \draw[thick] (-1,-1) -- (2,2);
        
        \node[anchor = center] at (1.3,1.7) {\rom{1}};
        \node[anchor = center] at (1.7,1.3) {\rom{2}};
        \node[anchor = center] at (0.5,1.5) {\rom{3}};
        \node[anchor = center] at (1.5,0.5) {\rom{4}};
        \node[anchor = center] at (-0.5,1.5) {\rom{5}};
        \node[anchor = center] at (1.5,-0.5) {\rom{6}};
        \node[anchor = center] at (0.3,0.7) {\rom{7}};
        \node[anchor = center] at (0.7,0.3) {\rom{8}};
        \node[anchor = center] at (-0.5,0.5) {\rom{9}};
        \node[anchor = center] at (0.5,-0.5) {\rom{10}};
        \node[anchor = center] at (-0.7,-0.3) {\rom{11}};
        \node[anchor = center] at (-0.3,-0.7) {\rom{12}};

        \node[anchor = center] at (2.1,1) {$1$};
        \node[anchor = center] at (1,2.125) {$1$};
    \end{tikzpicture}
    \caption{The different kinematic regions in  \((\tilde{\cQ}_{2,2},\tilde{\cQ}_{3,3})\)-space relevant for the triangle integral.}
    \label{fig:analytic_regions}
\end{figure}

We now discuss how many different kinematic regions we need to consider.
It is convenient to normalize the Gram matrix entries \(\cQ_{i,i}\) by \(\cQ_{1,1}=m_k>0\), which is always a mass and thereby positive. We define
\beq
\tilde{\cQ}_{i,i} = \frac{\cQ_{i,i}}{\cQ_{1,1}}\,.
\eeq
This leaves us with \(N-1\) independent quantities \(\tilde{\cQ}_{2,2}, \ldots, \tilde{\cQ}_{N,N}\).
We then need to consider all orderings of the set of \(N+1\) elements \(\{0,1,\tilde{\cQ}_{2,2}, \ldots, \tilde{\cQ}_{N,N}\}\), subject to the constraint \(0<1\). Hence the total number of accessible rearrangements is \(\tfrac{(N+1)!}{2}\).
For the triangle integral ($N=3$), this leads to twelve distinct regions, shown in figure~\ref{fig:analytic_regions}.
%However, s
Since the results for different orthoschemes are related by a simple relabeling of the variables, we will only discuss the orthoscheme defined by the ordering \(\mathbf{j}=(1,\ldots,N)\) in the following.
%
%The goal of this section is to describe a general strategy for analytically continuing our results from Euclidean kinematics to all other kinematic regimes. We demonstrate this systematic procedure for the case of the triangle diagram.

Let us now discuss the analytic continuation of this orthoscheme for $N=3$ to the twelve kinematic regions shown in figure~\ref{fig:analytic_regions}.
Each orthoscheme depends on the two dimensionless variables $x$ and $y$ defined in eq.~\eqref{eq:xy_Tri}. They involve square roots, and we fix their branches by assigning a small positive imaginary part:
\begin{align}
    x= \sqrt{\frac{\cQ_{2,2}-\cQ_{1,1}}{\cQ_{2,2}} + \iu 0} \textnormal{~~~and~~~}  y= \sqrt{\frac{\cQ_{2,2}-\cQ_{3,3}}{\cQ_{2,2}}+ \iu 0}\,.
\end{align}
We adopt the same convention for the square root in the normalisation of the integral, i.e.,
\begin{equation}
    I^{D}_N(Q) = \frac{1}{2}\left(\frac{-\iu}{2}\right)^{\tfrac{d}{2}}\, \frac{J^{D}_N(Q)}{\sqrt{\det Q^{(N)}+\iu 0}}\,.
\end{equation}
In order to simplify the notation, we drop the explicit $i0$ factors from the expressions.
%In the following we will assume \(x\) and \(y\), and the determinant factor \(\sqrt{\det Q^{(N)}}\) to have non-negative real and imaginary part and drop the \(+\iu 0^+\) form the expressions for brevity.

%Next, let us discuss the location of the boundaries between different kinematics regions. 
%The Gram matrix entries \(\cQ_{i,i}\) can be reformulated in terms of kinematic invariants \(\{m_i\}_{i=1,2,3}\) and \(\{p_i^2\}_{i=1,2,3}\) using eqs.~\eqref{eq:mathcalQ_coords} and \eqref{eq:Cayley}:
%\begin{align}
%    \Q_{1,1} &= m_1^2 \label{eq:tri_QQ11}\,,\\
%    \Q_{2,2} &= \frac{(p_1^2+(m_1-m_2)^2)(p_1^2+(m_1+m_2)^2)}{4 p_1^2}\label{eq:tri_QQ22}\,,\\
%    \Q_{3,3} &= \frac{ 4 \det Q }{\lambda (p_1^2,p_2^2,p_3^2)}\label{eq:tri_QQ33}\,,
%\end{align}
%with the Käll\'en function \(\lambda (a,b,c) = a^2+b^2+c^2-2a b-2 ac -2 b c\).
%We may insert these expression into the definitions of $x$ and $y$ to obtain:
%\begin{align}
    %x &= \sqrt{\frac{(p_1^2-m_1^2+m_2^2)^2}{(p_1^2+(m_1-m_2)^2)(p_1^2+(m_1+m_2)^2)}}\,,\label{eq:x_kinematic_var}\\[1.25ex]
%    y &=\sqrt{ - \frac{\left( p_1^2 (m_1^2 +m_2^2 - 2m_3^2 +p_1^2-p_2^2 - p_3^2) + (m_1^2-m_2^2)(p_2^2-p_3^2) \right)^2}{(p_1^2+(m_1-m_2)^2)(p_1^2+(m_1+m_2)^2)(-\lambda(p_1^2,p_2^2,p_3^2))} }\,.\label{eq:y_kinematic_var}
%\end{align}
%From  eq.~\eqref{eq:x_kinematic_var} and \eqref{eq:y_kinematic_var} we see that at the pseudo- and physical threshold \(-p_1^2 = (m_1 \mp m_2)^2\) the structure of \(x\) and \(y\) changes, which is to be expected since the analytic structure of \(I_3^4(Q)\) changes across these thresholds.

We now discuss the analytic continuation of an orthoscheme appearing in the projective dissection of the triangle integral to different kinematic regions.
We start from the Euclidean region \(0<1<\tilde{\cQ}_{2,2} < \tilde{\cQ}_{3,3}\), indicated by \rom{1} in figure~ \ref{fig:analytic_regions}. We discuss in detail the continuation to the regions \rom{2}, \rom{3} and \rom{5}. All other cases can be treated similarly. It is convenient to work directly with expressions that are antisymmetrized in $x$ and $y$. However, we will present expressions before acting with the antisymmetrizer $A_{xy}$ defined in eq.~\eqref{eq:A_xy_def}, because they are shorter. Note that the sign function and the absolute value are artifacts of working with the non-antisymmetrized expressions, and they can be removed after antisymmetrization. In particular, we have the relations \(A_{xy} \sgn(x) f(y) = 2 A_y f(y)\) and \(A_{xy} \sgn(xy) = 4\). 

Let us start by discussing the analytic continuation of the leading term in the $\eps$-expansion.
We will indicate regions with a superscript and the associated Roman numeral in brackets. The result in Region \rom{1} (the Euclidean region) can be found in eq.~\eqref{eq:tri_eps=0}. % given by
%\beq\label{eq:Region_I_result}
%    \hat{J}^{4,(\textrm{\rom{1}})}_{3,0} = [G(1;x)-G(1;y)]\, G(y;x)-G(1,y;x)\,,
%\eeq
%as we recall from \eqref{eq:tri_eps=0}.
We now discuss the continuation to the regions \rom{2}, \rom{3} and \rom{5}. Note that in Region \rom{1} we have $0<x<1$ and $y$ is purely imaginary (cf.~eq.~\eqref{eq:xy_Tri}).

\paragraph{\rom{1} \(\to\) \rom{2}:} The variable \(x\) still fulfils \(0<x<1\), but \(y\) becomes real, because \(\cQ_{2,2}-\cQ_{3,3} > 0\) in the numerator of the square root. Furthermore, since \(\Q_{3,3}>\cQ_{1,1}\), we conclude that \(0<y<x<1\) in Region \rom{2}. 
Hence both \(G(y;x)\) and \(G(1,y;x)\) develop imaginary parts in this region. However, we expect that the triangle integral is real in Region \rom{2}, because it corresponds to a kinematics below threshold (this can also be confirmed by numerical calculations). To make this explicit, we bring \(\hat{J}^{4,(\textrm{\rom{2}})}_{3,0}\) into a form where $y$ appears as the upper integration limit. We find:
\beq\bsp
    \hat{J}^{4,(\textrm{\rom{2}})}_{3,0} &= -G(1;x) G(1;y)+[G(1;x) -G(1;y) ]\,G(x;y)\\
    &\qquad+G(1,x;y)+G(0,1;x)+G(0,1;y)\,.
\esp\eeq
This expression is manifestly real and agrees with the numerical result obtained with \texttt{pySecDec}.
\paragraph{\rom{1} \(\to\) \rom{3}:} The variable \(x\) is imaginary with \(0<|x|<|y|\). Since both variables have a non-zero imaginary part and $|x|<|y|$, the analytic expression remains essentially unchanged with respect to Region \rom{1}. More concretely, we find 
\beq\hat{J}^{4,(\textrm{\rom{3}})}_{3,0} =  - \hat{J}^{4,(\textrm{\rom{1}})}_{3,0}\,,
\eeq
where the additional minus sign comes from our conventions for \(x\) and \(y\), as well as the square root factor used in the normalisation of the result.
\paragraph{\rom{3} \(\to\) \rom{5}:} Region \rom{5} corresponds to a kinematic situation above threshold, so we expect a non-zero imaginary part to develop. In Region \rom{5}, the variables $x$ and $y$ are
\beq\bsp
    x &= \sqrt{\frac{-|\cQ_{2,2}|-\cQ_{1,1}}{-|\cQ_{2,2}|}} = \sqrt{\frac{|\cQ_{2,2}|+\cQ_{1,1}}{|\cQ_{2,2}|}}>1\,,\\
    y &= \sqrt{\frac{-|\cQ_{2,2}|-\cQ_{3,3}}{-|\cQ_{2,2}|}} = \sqrt{\frac{|\cQ_{2,2}|+\cQ_{3,3}}{|\cQ_{2,2}|}}>1\,,
\esp\eeq
such that \(y>x>1\). the individual MPLs in eq.~\eqref{eq:tri_eps=0} develop an imaginary part. Our goal is to rewrite the MPLs in eq.~\eqref{eq:tri_eps=0} in such a way that individual MPLs are real. To this effect, we introduce the auxiliary variables
\beq
x'=\frac{1}{x} \textrm{~~~and~~~} y'=\frac{1}{y}\,,
\eeq
and we have $0<y'<x'<1$. We then pass to a fibration basis for the MPLs where $y'$ appears as the upper integration variable (this can be achieved in an automated way using, e.g., {\sc PolyLogTools}).
The signs of the imaginary parts need to be carefully fixed as presribed by the $i0$ prescription for the Mandelstam invariants. We do this here by comparing the analytic results at some selected numerical points to {\tt PySecDec}.  We find %\claudecomment{Can we show the result with $x'$ and $y'$ in the fibration basis?}
%\beq\bsp
%    \hat{J}^{4,(\textrm{\rom{5}})}_{3,0} &= (G(0;|y|)-G(0;|x|)) G\left(\frac{y}{x};1\right)-G(x;1) G\left(\frac{y}{x};1\right)+G(y;1) G\left(\frac{y}{x};1\right)\\
%    &\qquad +G(x;1) G(y;1)+G\left(0,\frac{y}{x};1\right)-G\left(y,\frac{y}{x};1\right) +\frac{1}{2} \iu \pi ~  \sgn(x) G(y;1)\,.
%\esp\eeq
\beq\bsp
    \hat{J}^{4,(\textrm{\rom{5}})}_{3,0} &= (G(0;|x'|)-G(0;|y'|)) G\left(x';y'\right)-G(1;x') G\left(x';y'\right)+G(1;y') G\left(x';y'\right)\\
    &\qquad +G(1;x') G(1;y')+G\left(0,x';y'\right)-G\left(1,x';y'\right) +\frac{1}{2} \iu \pi ~  \sgn(x') G(1;y')\,.
\esp\eeq

\begin{figure}[h!]
    \centering
    \includegraphics[height=0.35\linewidth]{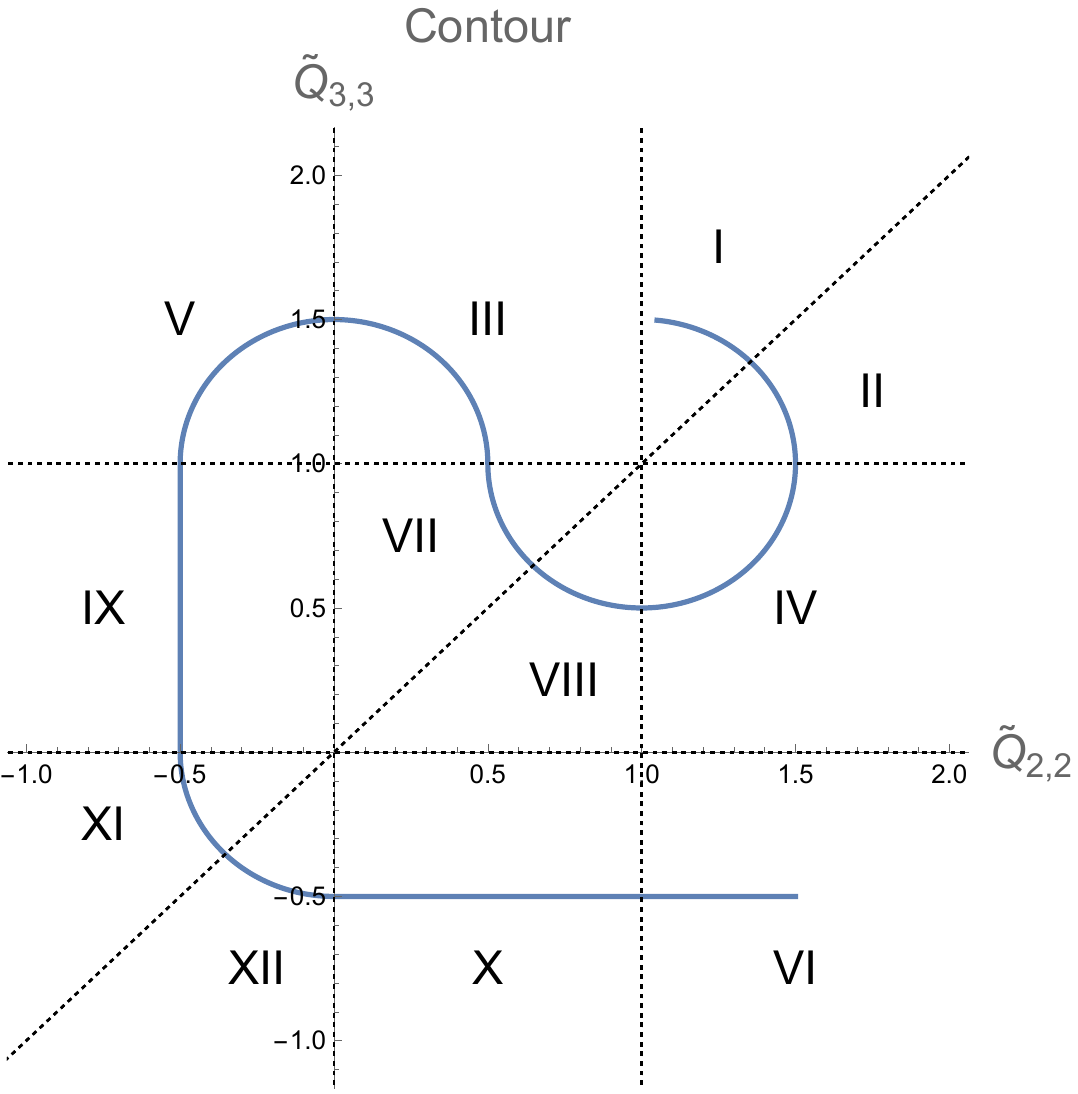}
    \includegraphics[height=0.35\linewidth]{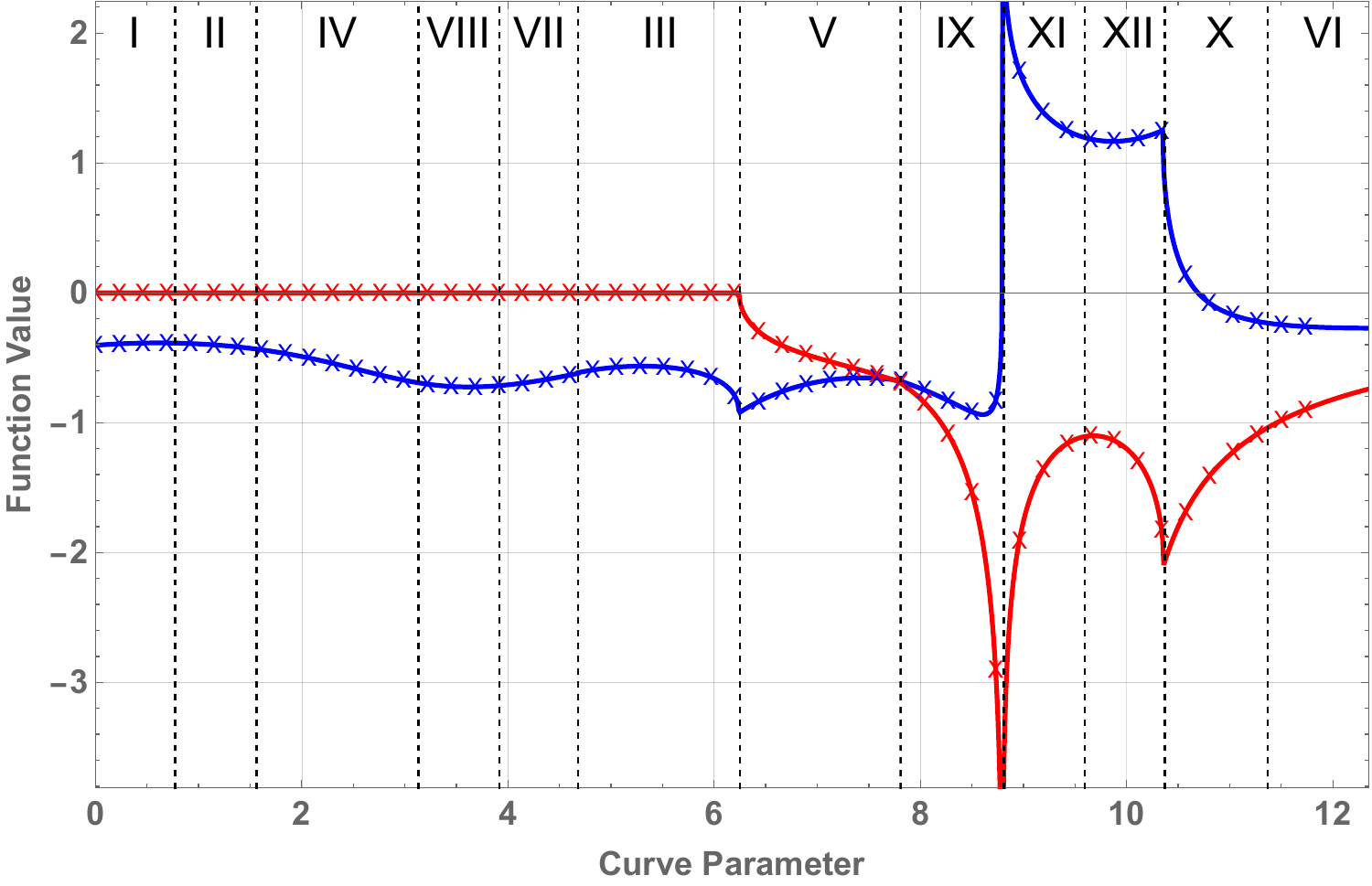}
    \caption{(left) Contour $\gamma(t)=(\tilde{\cQ}_{2,2}(t),\tilde{\cQ}_{3,3}(t))$ in the normalized $\tilde{\cQ}_{2,2}$--$\tilde{\cQ}_{3,3}$ plane used for analytic continuation. 
    (right) Value of $I_3^4(\cQ_{1,1},\cQ_{1,1}\tilde{\cQ}_{2,2}(t),\cQ_{1,1}\tilde{\cQ}_{3,3}(t))$ along the contour, plotted versus the curve parameter $t$. The solid lines are obtained from our analytic result (real: blue, imaginary: red), while the crosses represent numerical results from \texttt{pySecDec}.}
    \label{fig:analytic_continuation_tri_eps=0}
\end{figure}

\paragraph{Summary of the results for the different regions.} The method illustrated on Regions \rom{1}, \rom{2}, \rom{3} and \rom{5} is exemplary, and can be applied to all other regions. We find:
\begin{enumerate}
  \renewcommand{\labelenumi}{\Roman{enumi}.} % Uppercase Roman numerals
  \item \(\hat{J}^{4,(\textrm{\rom{1}})}_{3,0} = [G(1;x)-G(1;y)] G(y;x)-G(1,y;x)\,\),
  \item \(\hat{J}^{4,(\textrm{\rom{2}})}_{3,0} = -G(1;x) G(1;y)+[G(1;x) -G(1;y) ]G(x;y)+G(1,x;y)+G(0,1;x)+G(0,1;y)\,\),
  \item \(\hat{J}^{4,(\textrm{\rom{3})}}_{3,0} =-\hat{J}^{4,(\textrm{\rom{1}})}_{3,0} \),
  \item \(\hat{J}^{4,(\textrm{\rom{4}})}_{3,0} =\hat{J}^{4,(\textrm{\rom{1}})}_{3,0} \),
  \item \(\hat{J}^{4,(\textrm{\rom{5}})}_{3,0} = (G(0;|x'|)-G(0;|y'|)) G\left(x';y'\right)-G(1;x') G\left(x';y'\right)+G(1;y') G\left(x';y'\right) \\ \phantom{\hat{J}^{4,(\textrm{\rom{5}})}_{3,0} =} +G(1;x') G(1;y')+G\left(0,x';y'\right)-G\left(1,x';y'\right) +\frac{1}{2} \iu \pi ~  \sgn(x') G(1;y')\,, \)
  %\( \hat{J}^{4,(\textrm{\rom{5}})}_{3,0} = [G(0;|y|)-G(0;|x|)]\,G\left(\tfrac{y}{x};1\right)-G(x;1) G\left(\frac{y}{x};1\right)+G(y;1) G\left(\frac{y}{x};1\right) +G(x;1) G(y;1)+G\left(0,\frac{y}{x};1\right)-G\left(y,\frac{y}{x};1\right) +\frac{1}{2} \iu \pi ~  \sgn(x) G(y;1)\,\),
  \item \(\hat{J}^{4,(\textrm{\rom{6}})}_{3,0} = -G(1,y;x)+G(y;x) \left[G(1;x)-G(y;1)-G(0;|y|)\right]+\frac{i \pi }{2} ~G(y;x)\,\),
  \item \(\hat{J}^{4,(\textrm{\rom{7}})}_{3,0} = - \hat{J}^{4,(\textrm{\rom{2}})}_{3,0}\),
  \item \(\hat{J}^{4,(\textrm{\rom{8}})}_{3,0} = \hat{J}^{4,(\textrm{\rom{2}})}_{3,0}\),
  \item \(\hat{J}^{4,(\textrm{\rom{9}})}_{3,0} = (G(0;|x'|)-G(0;|y'|)) G\left(y';x'\right)-G(1;x') G\left(y';x'\right)+G(1;y') G\left(y';x'\right) \\
  \phantom{\hat{J}^{4,(\textrm{\rom{9}})}_{3,0} =} -G\left(0,y';x'\right)+G\left(1,y';x'\right)-\frac{1}{4} \pi ^2 \sgn(x' y')   +\frac{1}{2} \iu \pi ~ \sgn(x') G(1;y')\,,\)
  %\(\hat{J}^{4,(\textrm{\rom{9}})}_{3,0} = (G(0;|y|)-G(0;|x|)) G\left(\frac{x}{y};1\right)-G(x;1) G\left(\frac{x}{y};1\right)+G(y;1) G\left(\frac{x}{y};1\right)-G\left(0,\frac{x}{y};1\right)+G\left(x,\frac{x}{y};1\right)-\frac{1}{4} \pi ^2 \sgn(x y)   +\frac{1}{2} \iu \pi ~ \sgn(x) G(y;1),\),
  \item \(\hat{J}^{4,(\textrm{\rom{10}})}_{3,0} = \hat{J}^{4,(\textrm{\rom{6}})}_{3,0}\),
  \item \(\hat{J}^{4,(\textrm{\rom{11}})}_{3,0}  = -G(0;|x|) G(x;y)+G\left(1;\frac{1}{x}\right) G(1;y)+G(1;y) G(x;y)-G\left(1;\frac{1}{x}\right) G(x;y) \\ \phantom{\hat{J}^{4,(\textrm{\rom{11}})}_{3,0}  =} -G(1,x;y) +\frac{1}{2} \iu \pi ~ \left( \sgn(x) G(1;y) - G(x;y)\right)\,\),
  \item \(\hat{J}^{4,(\textrm{\rom{12}})}_{3,0} = G(0;|x'|)  G\left(\tfrac{1}{y};x'\right)-G(1;x') G\left(\frac{1}{y};x'\right)+G\left(1;y\right) G\left(\frac{1}{y};x'\right)-G\left(0,\frac{1}{y};x'\right) \\ \phantom{\hat{J}^{4,(\textrm{\rom{12}})}_{3,0} =} +G\left(1,\frac{1}{y};x'\right)    +\frac{1}{2} \iu \pi ~ \left( \sgn(x') G(1;y) -   G\left(\frac{1}{y};x'\right)  \right)\,\). 
  %\(\hat{J}^{4,(\textrm{\rom{12}})}_{3,0} = -G(0;|x|)  G(x;y)-G(x;1) G\left(\frac{x}{y};1\right)+G\left(\frac{1}{y};1\right) G\left(\frac{x}{y};1\right)-G\left(0,\frac{x}{y};1\right)+G\left(x,\frac{x}{y};1\right)    +\frac{1}{2} \iu \pi ~ \left( \sgn(x) G(1;y) -   G\left(\frac{x}{y};1\right)  \right)\,\).
\end{enumerate}
These expressions are sufficient to obtain numerical results everywhere in the kinematic space for the triangle integral in exactly $D=4$ dimensions. 
We have checked that our result is correct by comparing it to a direct numerical evaluation with {\tt pySecDec} for a set of kinematic points chosen on the one-dimensional slice in parameter space shown in the left panel in figure~\ref{fig:analytic_continuation_tri_eps=0}.
In the right panel we can see that we find perfect agreement with {\tt pySecDec} everywhere.

%We validated our results, which were fixed through a single sampling point in each region, by comparing the full result against many more sampling points, obtained through numerical calculations with \texttt{pySecDec}, and found very good agreement. The results for \(\eps^0\) and \(\eps^1\) can be found in fig.~\ref{fig:analytic_continuation_tri_eps=0} and \ref{fig:analytic_continuation_tri_eps1} respectively.

\begin{figure}[h!]
    \centering
    \includegraphics[height=0.35\linewidth]{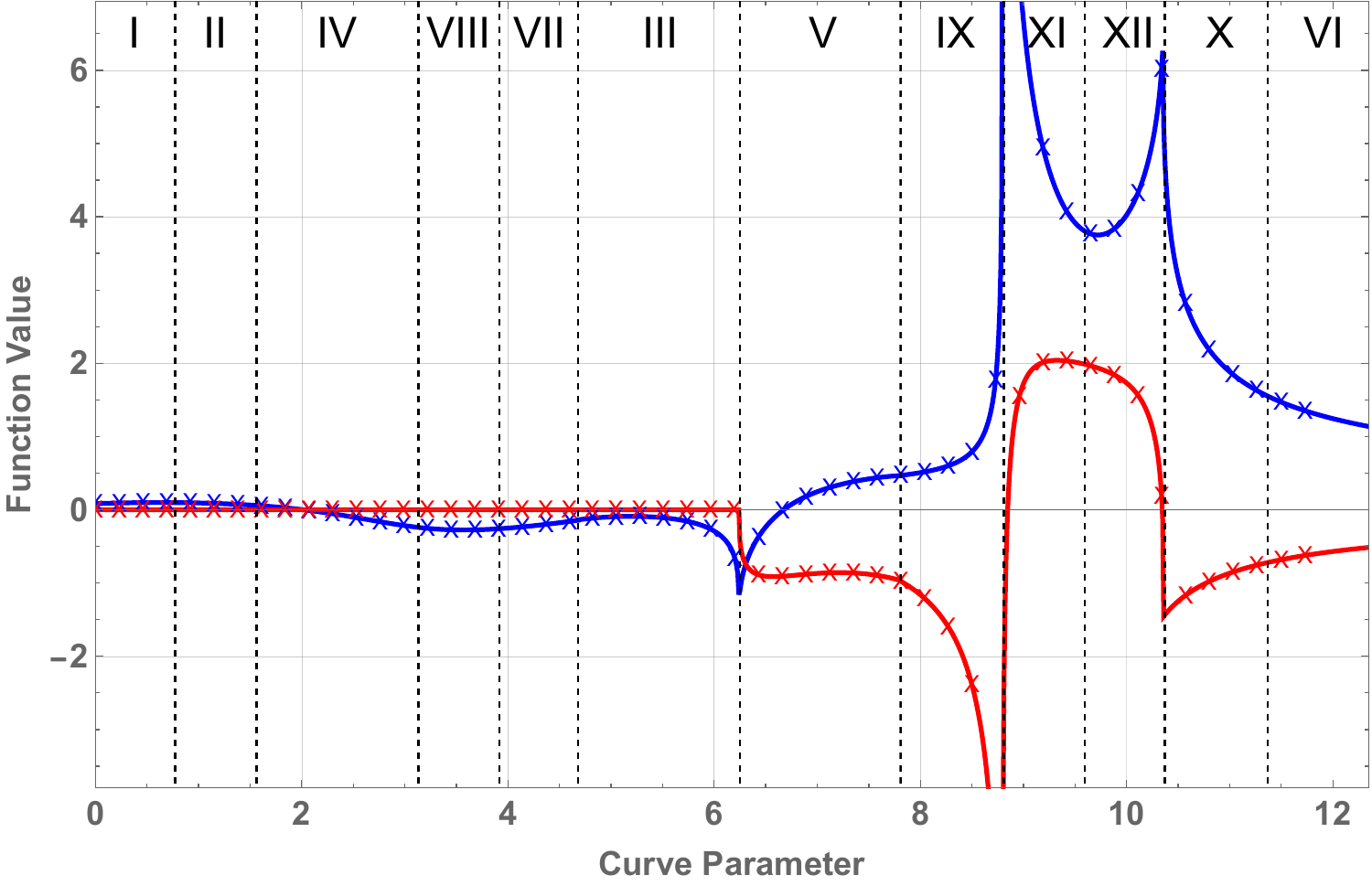}
    \caption{
    %\paulcomment{@paul check if this is without EulerGamma already} 
    Comparison of the \(\mathcal{O}(\eps^1)\) contribution from \(I_3^{4-2\eps}(\cQ(t))\) with the numerical results from \texttt{pySecDec} using the same conventions as in figure~\ref{fig:analytic_continuation_tri_eps=0}.}
    \label{fig:analytic_continuation_tri_eps1}
\end{figure}

So far we have only discussed the analytic continuation for $\eps=0$. There is no conceptual obstacle to extend our strategy to higher orders in the $\eps$-expansion.
In particular, we have also obtained the analytic continuation for the $\mathcal{O}(\eps^1)$ terms. In figure~\ref{fig:analytic_continuation_tri_eps1} we show that we find perfect agreement with {\tt pySecDec} in all regions.
The analytic expressions for the first two orders in the $\eps$-expansion of the triangle integral can be found in the ancillary file \texttt{results.wl}. %\paulcomment{@paul need to add to file}. 

Let us conclude this section by a comment.
There are actually two ways to compute the analytic continuation for higher orders in \(\eps\). The first one is the strategy discussed in this section, and consists in using our analytic results for the higher orders in the expansion in the Euclidean region and to continue them to other regions.
The second one consists in performing the analytic continuation only on the $\mathcal{O}(\eps^0)$ contribution, and then to perform the auxiliary-mass integration region by region on the continued expressions. We find the first approach to be simpler to implement.

\section{Conclusions}
\label{sec:conclusions}

In this paper, we have introduced an algorithm to compute the higher orders in the dimensional regulator $\eps$ of one-loop triangle, box and pentagon integrals depending on an arbitrary number of scales in a `t Hooft-Veltman-like scheme where all external momenta lie in four space-time dimensions. Our algorithm heavily relies on two technical advances in both mathematics and physics. On the one hand, we rely on the fact that one-loop integrals for $\eps=0$ compute volumes of simplices in hyperbolic spaces. Very recently, a general algorithm was introduced to evaluate such volumes in terms of MPLs~\cite{Rudenko2020Orthoschemes}. This allows one to obtain analytic formulas for all one-loop integrals at $\eps=0$~\cite{Ren:2023tuj}. On the other hand, we use modern advances in our understanding of how to evaluate integrals involving MPLs using direct integration~\cite{Brown2009MasslessHigherLoop,Anastasiou:2013srw,Ablinger:2014yaa,Panzer:2014caa,hyperlogprocedures,Kardos:2025klp}. Remarkably, we find that for up to pentagon integrals, these two results can be seamlessly combined. In particular, all square roots can be rationalised, and all integrations can be performed analytically in terms of MPLs.

We have illustrated our algorithm by computing the first orders in the $\eps$-expansion of triangle, box and pentagon integrals depending on arbitrary propagator and external masses. We stress that these computations are exemplary, and we expect the same strategy to be applicable to the higher orders in the $\eps$-expansion or for integrals depending on fewer scales. In particular, our results provide a rigorous proof that all triangle, box and pentagon integrals can be expressed in terms of MPLs to all orders in dimensional regularisation. 

For the future it would be interesting to investigate if it is possible to exploit the connection between one-loop integrals and volumes of simplices in hyperbolic spaces also for integrals in dimensional regularisation depending on six or more external legs. This is currently under investigation~\cite{Morkinprep}.

\acknowledgments
The authors are grateful to Babis Anastasiou for discussions that led to the present work, and to Steven Charlton and Herbert Gangl for discussions. All figures of orthoschemes (simplices) have be produced with \texttt{GeoGebra}. This work is funded in part by the European Union
(ERC Consolidator Grant LoCoMotive 101043686). Views
and opinions expressed are however those of the author(s)
only and do not necessarily reflect those of the European
Union or the European Research Council. Neither the
European Union nor the granting authority can be held
responsible for them. 

%Appendices
\appendix
\renewcommand{\appendixname}{}

%\section{Some proofs omitted in the main text}\label{app:proofs}
%\paulcomment{in progress}
\section{Additivity in the auxiliary mass parameter of the orthoscheme Gram matrix}
\label{app:proof_additivity}

In this appendix we present the proof of eq.~\eqref{eq:cQ+eta}. The proof is a simple application of the matrix determinant lemma and Schur's formula.

Assume \(A\) to be a symmetric and invertible matrix. If $u$ and $v$ are vectors, the matrix determinant lemma states
\beq
    \det (A + u v^\top) = \left(1 + v^\top A^{-1} u \right) \det A\,.
\eeq
With the choice $u=\eta\,\mathbf{1}$ and $v=\mathbf{1}$, we obtain:
\beq\label{eq:A+eta11}
    \det(A+\eta \, \mathbf{1} \mathbf{1}^\top ) = (1 + \eta\, \mathbf{1}^\top A^{-1} \mathbf{1}) \det A \,.
\eeq
Using the Schur's formula,
\beq
    \det \begin{pmatrix} A & B \\ C & D \end{pmatrix} = \det A \det (D-C A^{-1} B)\,,
\eeq
we can find a relation between \(\mathbf{1}^\top A^{-1} \mathbf{1}\) and the determinant of \(A_\textnormal{ext}\):
\beq\label{eq:1.A^-1.1__Aext_relation}
    \det A_\textnormal{ext} = \det \begin{pmatrix} A & \tfrac{1}{2} \mathbf{1}\\ \tfrac{1}{2} \mathbf{1}^\top & 0 \end{pmatrix} = \det A \det \left( -\tfrac{1}{4} \mathbf{1}^\top A^{-1} \mathbf{1} \right) = \left( -\tfrac{1}{4} \mathbf{1}^\top A^{-1} \mathbf{1} \right) \det A \,.
\eeq
Combining eqs.~\eqref{eq:A+eta11} and \eqref{eq:1.A^-1.1__Aext_relation}, we can deduce that 
\beq
    \det(A+\eta \, \mathbf{1} \mathbf{1}^\top )  = \det A ~ \left(1 +\tfrac{- 4 \det A_\textnormal{ext}}{ \det A} \, \eta  \right)\,.
\eeq
On the other hand, as consequence of the multilineality of the determinant, we have
\beq
    \det\left( (A+\eta \, \mathbf{1} \mathbf{1}^\top)_\textnormal{ext} \right) = \det \begin{pmatrix} A+\eta \, \mathbf{1} \mathbf{1}^\top & \tfrac{1}{2} \mathbf{1}\\ \tfrac{1}{2} \mathbf{1}^\top & 0 \end{pmatrix} = \det A_\textnormal{ext}\,.
\eeq
Hence 
\beq
    \frac{\det(A+\eta \, \mathbf{1} \mathbf{1}^\top )}{-4\det\left( (A+\eta \, \mathbf{1} \mathbf{1}^\top)_\textnormal{ext} \right) } = \frac{\det A}{-4 \det A_\textnormal{ext}} + \eta\,.
\eeq
This proves our claim.  

%We deduce that
%\beq
%    \mathbf{1}^\top A^{-1} \mathbf{1} = -4 \frac{\det A_\textnormal{ext}}{\det A}\,.
%\eeq
%This relation is a
%\beq
%    \det (A + u v^\top) = \det A \left(1 + v^\top A^{-1} u \right)
%\eeq
%to calculate:
%\beq
%    \det(A+\eta \, \mathbf{1} \mathbf{1}^\top ) = \det A (1 + \eta\, \mathbf{1}^\top A^{-1} \mathbf{1}) = \det A ~ \left(1 +\tfrac{- 4 \det %A_\textnormal{ext}}{ \det A} \, \eta  \right).
%\eeq

\section{Proof of \(\operatorname{Dis}_{P_\infty}(Q) = \operatorname{BS} (Q)\)}
\label{eq:proj_split_equiv}
%\paulcomment{@paul need to be careful with signs of momenta here check}
In this appendix we prove the equivalence of the splitting of the basic simplex and the projective dissection. More precisely, we will show that 
\beq
\operatorname{Dis}_{P_\infty}(Q) = \operatorname{BS} (Q)\,,
\eeq
where $\operatorname{Dis}_{P_\infty}(Q)$ and $\operatorname{BS} (Q)$ denote the set of Gram matrices for the two decompositions, respectively.
The goal of this section is to show that \(\operatorname{Dis}_{P_\infty}(Q) = \operatorname{BS} (Q)\). Let us pick the Gram matrix $\cQ(\mathbf{j})\in\operatorname{Dis}_{P_\infty}(Q)$ that arises from the projective dissection. We need to show that $\cQ(\mathbf{j})$ also lies in $\operatorname{BS}(Q)$. Since $\operatorname{Dis}_{P_\infty}(Q)$ and  $\operatorname{BS} (Q)$ have the same number of elements, it follows that they must be the same. It is sufficient to do this for the Gram matrix $\cQ(\mathbf{j})$ for the standard ordering \(\mathbf{j}=(1,2,3,\ldots,N)\). All other orderings can be shown in the same way. For readability, we drop the dependence on $\mathbf{j}$.

% By applying the considerations of ref.~\cite{Ren:2023tuj} to the extended Gram matrix \(Q_\textnormal{ext}\) we already got eq.~\eqref{eq:mathcalQ_coords} for the entries in the Gram matrix \(\cQ(\mathbf{j})\). Therefore, we only need to derive the form of \(\cQ (\mathbf{j})\) for the basic simplex splitting.

%We will only need look at one specific orthoscheme, i.e., we pick \(\mathbf{j}=(1,2,3,\ldots,N)\). All other orthoschemes follow analogously. For brevity we write \(\cQ = \cQ(1,2,\ldots,N)\), which by 

From eq.~\eqref{eq:mathcalQ_coords} we know that (without loss of generality we may pick \(i \le k\)):
\beq\label{eq:mathcalQ(123...)}
    \cQ_{i,k} = \cQ_{k,k} = \frac{\det Q_{\{1,2,\ldots, k\}}}{-4 \det Q_{\{1,2,\ldots, k\}}^\textnormal{ext}}\,,
\eeq
where \(Q_{\{1,2,\ldots, k\}}\) is the \(k \times k\) submatrix given by considering the first \(k\) rows and columns and \(Q_{\{1,2,\ldots, k\}}^\textnormal{ext} = (Q_{\{1,2,\ldots, k\}})_\textnormal{ext}\) is the extended matrix associated with this submatrix \(Q_{\{1,2,\ldots, k\}}\) (cf.~eq.~\eqref{eq:Qext_def}).

\begin{figure}[!th]
    \centering
    \begin{tikzpicture}[scale=1.1,>=Latex, every node/.style={font=\small}]
        
        %--- vertices ---------------------------------------------------------
        \coordinate (M)  at (0,0);        % left apex
        \coordinate (A)  at (5.5,-0.5);     % base vertex 1
        \coordinate (B)  at (7, 1.2);     % base vertex 2
        \coordinate (C)  at (5.2,3.0);    % base vertex 3
        \coordinate (G)  at (5.9,1.3); % interior point (near centroid)
        \coordinate (Gp)  at (6.25,0.35); % interior point (near centroid)
        
        %--- black edges from M -----------------------------------------------
        \draw[black] (M) -- (A);
        \draw[red] (M) -- (A) node[midway,below=2pt] {$m_1$};
        \draw[black] (M) -- (B) node[midway, xshift=35pt, yshift=-2pt] {$m_2$};
        \draw[black] (M) -- (C) node[midway,above left=-1pt] {$m_3$};
        
        %--- red base triangle and inner connectors ---------------------------
        \draw[black] (A) -- (B); %node[midway,below=2pt] {$L_{12}$};
        \draw[black] (A) -- (C) node[midway,above left=2pt] {$L_{13}$};
        \draw[black] (B) -- (C) node[midway,right=2pt] {$L_{23}$};
        
        % optional interior red connectors (like your sketch)
        \draw[red] (M) -- (G) node[midway,above left=-1pt] {$m_{123}$};
        \draw[red] (A) -- (G);
        \draw[red] (G) -- (Gp);
        \draw[red] (A) -- (Gp) node[midway,below right=-2pt] {$L'_{12}$};
        \draw[red,dashed] (M) -- (Gp) node[midway,above right=-2pt] {$m_{12}$};
        
        %--- nodes -------------------------------------------------------------
        \fill[red] (M) circle (1.2pt) node[left] {$M$};
        \fill[red] (A) circle (1.2pt) node[below] {$v_1$};
        \fill (B) circle (1.2pt) node[right] {$v_2$};
        \fill (C) circle (1.2pt) node[above] {$v_3$};
        \fill[red] (G) circle (1.2pt) node[right] {$v_{123}$};
        \fill[red] (Gp) circle (1.2pt) node[right] {$v_{12}$};

        % Misc
        \pic [draw, angle radius=6pt, thick, red] {right angle = M--G--Gp};
        \pic [draw, angle radius=6pt, thick ,red] {right angle = G--Gp--A};
    \end{tikzpicture}
    \caption{Schematic depiction of the basic simplex for a triangle diagram with one of six orthoschemes marked in red.}
    \label{fig:basic_simplex_tri}
\end{figure}

We need to show that the same Gram matrix also arises from splitting the basic simplex. Recall that the basic simplex is defined by means of  Euclidean geometry (see section~\ref{sec:basic_simplex}). We refer to the new vertices on the hyperplane defined by the vectors \(\{\mathbf{m}_i,\mathbf{m}_j,\mathbf{m}_k,\ldots\}\) by $v_{ijk\ldots}$ (see figure~\ref{fig:basic_simplex_tri}).
The corresponding new vectors connecting the mass meeting point \(M\) to these vertices are denoted by their vertex labels. For example, \(\mathbf{m}_{12\ldots k}\) is the vector from \(M\) to \(v_{12\ldots k}\). Likewise, edges between two hypersurface vertices \(v_{12\ldots i}\) and \(v_{12\ldots j}\) (with \(i<j\)) are denoted by the momentum length $
  L'_{ij}=\sqrt{-(l'_{ij})^2}$.
Since we are in Euclidean geometry, we can use the Pythagorean theorem to rewrite the value of \(L'_{ij}\) in terms of the heights \(m_{12\ldots k}\):
\beq
    (L'_{ij})^2 = m_{12\ldots j}^2 - m_{12\ldots i}^2 \,.
\eeq
This results in the following Gram matrix:
\beq
    (\cQ^{\mathrm{BS}})_{ij} = \frac{m_{12\ldots i}^2 + m_{12\ldots j}^2 - (L'_{ij})^2}{2} = m_{1\ldots \max(i,j)}^2\,, \qquad i,j=1,\ldots,N\,,
\eeq
or more explicitly:
\beq\label{eq:cQ^BS}
    \cQ^\textnormal{BS} = 
    \begin{pmatrix}
        m_{1}^2            & m_{12}^2         & m_{123}^2          & \cdots & m_{12\ldots N}^2 \\
        m_{12}^2           & m_{12}^2          & m_{123}^2          & \cdots & \vdots          \\
        m_{123}^2          & m_{123}^2         & m_{123}^2          & \cdots & \vdots          \\
        \vdots           & \vdots          & \vdots           & \ddots & \vdots          \\
        m_{12\ldots N}^2   & \cdots          & \cdots           & \cdots & m_{12\ldots N}^2
    \end{pmatrix}\,.
\eeq
We can already observe the form of the Gram matrices in the basic simplex picture and the projective dissection (cf.~eq.~\eqref{eq:Gram_special}).
The \(m_{12\ldots k}\) are orthogonal heights and can be computed by fundamental geometric means. One has the formula:
\begin{equation}
    m_{12\ldots k}^2 = \frac{\det Q_{\{1,2,\ldots,k\}}}{\det(l_{i,*} \cdot l_{j,*})_{i,j\in \{1,2,\ldots,k \}\setminus \{*\}}} = \frac{\det Q_{\{1,2,\ldots,k\}}}{-4 \det Q_{\{1,2,\ldots,k\}}^\textnormal{ext}}\,,
\end{equation}
where \((l_{i,*} \cdot l_{j,*})\) is the Gram matrix from vectors \(l_{i,*} = l_i - l_*\) where \(*\) is an arbitrary index in \(\{1,\ldots,N\}\).
Note that eq.~\eqref{eq:cQ^BS} is the same as eq.~\eqref{eq:mathcalQ(123...)}, as expected. We deduce from symmetry that also \(\operatorname{Dis}_{P_\infty}(Q) = \operatorname{BS} (Q)\) holds.

\bibliographystyle{JHEP}
\bibliography{biblio,biblio2}

\end{document}